\documentclass[aps,prd,reprint,preprintnumbers,superscriptaddress,nofootinbib,longbibliography,floatfix]{revtex4-2}
\usepackage{amssymb}
\usepackage{amsthm}
\usepackage{amsmath}
\usepackage{multirow}

\usepackage{hhline}
\usepackage[font=small,labelfont=bf]{caption}
\usepackage{subcaption}
\usepackage{cancel}
\usepackage{url}

\usepackage{array}
\usepackage{mathtools}

\usepackage{hyperref}
\hypersetup{
  colorlinks=true,
  citecolor=blue,
  linkcolor=blue,
  urlcolor=blue
}

\newcommand{\Z}{\mathbb{Z}}

\newcommand{\R}{\mathbb{R}}

\newcommand{\E}{\mathbb{E}}

\newcommand{\mae}{\text{MAE}}
\newcommand{\lr}{\mathcal{L}}

\renewcommand{\d}[1]{\ensuremath{\operatorname{d}\!{#1}}}

\DeclareMathOperator*{\maxi}{argmin}

\begin{document}

\title{Learning Likelihood Ratios with Neural Network Classifiers}

\author{Shahzar Rizvi}
\email{shahzar@berkeley.edu}
\affiliation{Department of Statistics, University of California, Berkeley, CA 94720, USA}

\author{Mariel Pettee}
\email{mpettee@lbl.gov}
\affiliation{Physics Division, Lawrence Berkeley National Laboratory, Berkeley, CA 94720, USA}

\author{Benjamin Nachman}
\email{bpnachman@lbl.gov}
\affiliation{Physics Division, Lawrence Berkeley National Laboratory, Berkeley, CA 94720, USA}
\affiliation{Berkeley Institute for Data Science, University of California, Berkeley, CA 94720, USA}

\begin{abstract}
    \vspace*{6pt}\noindent The likelihood ratio is a crucial quantity for statistical inference in science that enables hypothesis testing, construction of confidence intervals, reweighting of distributions, and more. Many modern scientific applications, however, make use of data- or simulation-driven models for which computing the likelihood ratio can be very difficult or even impossible. By applying the so-called ``likelihood ratio trick,'' approximations of the likelihood ratio may be computed using clever parametrizations of neural network-based classifiers. A number of different neural network setups can be defined to satisfy this procedure, each with varying performance in approximating the likelihood ratio when using finite training data. We present a series of empirical studies detailing the performance of several common loss functionals and parametrizations of the classifier output in approximating the likelihood ratio of two univariate and multivariate Gaussian distributions as well as simulated high-energy particle physics datasets. 
\end{abstract}

\maketitle

\vspace{\baselineskip}
    
 
\section{Introduction}
Claiming a scientific discovery requires a hypothesis test, i.e. a statistical threshold for claiming that one's experimental data reject the null hypothesis in favor of an alternative hypothesis. This might involve two probability densities:
\begin{itemize}
    \itemsep-0.55em 
    \item $H_0$ (the null hypothesis)
    \item $H_1$ (the alternative hypothesis)
\end{itemize}

By the Neyman-Pearson lemma \cite{neyman}, the strongest (``uniformly most powerful'') measure of whether the experimental data $x$ support $H_0$ vs. $H_1$ is a likelihood ratio test. These tests are particularly widespread in reporting results in High-Energy Physics (HEP), but are also commonly used for statistical analyses across astrophysics, biology, medicine, and other scientific domains concerned with hypothesis testing or confidence intervals. The need for likelihood ratios goes beyond hypothesis testing, too---they can also be used to reweight a distribution to align with a target distribution, such as reweighting simulation samples to match real data \cite{omnifold, Rogozhnikov_2016, Martschei_2012, Andreassen_2019, Andreassen_2020, Aaij_2017, PhysRevD.98.092002, fischer2023treating}.

In the simplest form of a likelihood ratio test, where $H_0$ and $H_1$ are fully-defined by parameters $\theta_0$ and $\theta_1$, the background-only hypothesis is either rejected (or not) depending on the value of the ratio of likelihoods $p(\theta_0 \mid x)$ under $H_0$ and $p(\theta_1 \mid x)$ under $H_1$ in relation to the desired significance level. 

In practice, however, the probability densities $H_0$ and $H_1$ may not be explicitly known. Worse, they might be nearly impossible to compute, such as in instances where they are generated by a complex simulation model. In these cases, we can use machine learning to directly approximate the likelihood ratio itself, bypassing the need to approximate the individual probability densities. 

A classifier function $f(x)$ (for instance, from a neural network) designed to distinguish data sampled from $H_0$ ($f(x) \rightarrow 0$) vs. $H_1$ ($f(x) \rightarrow 1$) can be used to approximate the likelihood ratio by minimizing a proper loss functional (defined in Section \ref{sec:learning}): 

\begin{equation}
\maxi_{f} L[f] = \frac{p(x \mid \theta_0)}{p(x \mid \theta_1)} = \mathcal{L}(x).
\end{equation}

For instance, in the familiar case of training a classifier by minimizing the binary cross-entropy loss (see \ref{tab:losses}), the optimal decision function $f(x)$ is:

\begin{equation}
    f(x) = \frac{p(x \mid \theta_0)}{p(x \mid \theta_0) + p(x \mid \theta_1)}.
\end{equation}

We can then approximate the likelihood ratio with a monotonic transformation of the neural network output $f(x)$\footnote{This notation assumes balanced training sets for simplicity. With imbalanced classes, one would need to modify the likelihood ratio to include prior factors $p(\theta_i)$, though the likelihood ratio trick will still apply \cite{Nachman_2021}.}:
\begin{align}
    \frac{f(x)}{1-f(x)} &= \frac{\frac{p(x \mid \theta_0)}{p(x \mid \theta_0) + p(x \mid \theta_1)}}{1 - \frac{p(x \mid \theta_0)}{p(x \mid \theta_0) + p(x \mid \theta_1)}} \\
    &= \frac{p(x \mid \theta_0)}{\cancel{p(x \mid \theta_0)} + p(x \mid \theta_1) - \cancel{p(x \mid \theta_0)}}\\
    &= \frac{p(x \mid \theta_0)}{p(x \mid \theta_1)} = \mathcal{L}(x).
\end{align}

This procedure, sometimes called the ``likelihood ratio trick'', is well-known in statistics (see e.g. \cite{hastie01statisticallearning,sugiyama_suzuki_kanamori_2012, Miller}) and has been frequently used in particle physics \cite{cranmer, Nachman_2020, Nachman_2021, Andreassen_2020, Andreassen_2021, hollingsworth2020resonance, Brehmer_2018, Brehmer_2018_long, brehmer2020madminer, badiali2020efficiency, salad, omnifold, Erdmann_2020, dagnolo2021learning,Diefenbacher:2020rna, hera}. 

A number of different loss functionals beyond binary cross-entropy can be defined to satisfy this setup, but in practice, not all such classifiers will perform equally well when approximating the likelihood ratio. In this paper, we perform a series of empirical studies to understand how different choices of loss functional and parametrization of the resulting classifier affect the performance of likelihood ratio approximation for pairs of distributions. 

Several recent works have investigated some improved configurations for the likelihood ratio trick in certain scientific contexts. \cite{kong} introduces a new likelihood estimation procedure as an extension of \cite{cranmer} using binary cross-entropy loss with SELU \cite{selu} activation. \cite{D_Agnolo_2019} notes that for one- and two-dimensional toy simulations of particle physics datasets, the maximum likelihood classifier (MLC) loss performed better than the binary cross-entropy loss when estimating the likelihood ratio---the first application of MLC loss in particle physics. \cite{Nachman_2021} directly compares linear and exponential parameterizations of maximum likelihood classifier loss with binary cross-entropy loss for one-dimensional Gaussians.
\cite{cranmer} uses calibrated classifiers to improve likelihood ratio estimation, and \cite{Brehmer_2018, Brehmer_2018_long} define several different approaches to likelihood ratio estimation, including augmenting the likelihood ratio trick with score regression (\textsc{Rascal}, \textsc{Sally}, etc.). \cite{cranmer2} introduces modified versions of the cross-entropy loss that show stronger performance under limited training dataset sizes than the typical cross-entropy loss, while  \cite{Moustakides} compares the estimation of the likelihood ratio via mean square loss with ELU \cite{elu} activation, cross-entropy loss with sigmoid activation, and a proposed exponential loss with no activation function on univariate Gaussian distributions. Still other methods use normalizing flows to determine the likelihood ratio by modeling the individual densities \cite{normalizingflow, anode} or to obviate the need for the likelihood ratio approximation for reweighting distributions \cite{algren2023flow}. \cite{jeffrey2023evidence} also proposes a novel \emph{l-POP-Exponential} loss that performs better than some traditional losses such as binary cross-entropy loss when estimating the log Bayes factor.

In light of these existing studies, this work serves as a detailed comparison of a wide range of configurations of loss functionals and output parametrizations across datasets including one-dimensional Gaussians, multi-dimensional Gaussians, and simulated high-energy particle physics datasets. We aim to highlight some best practices and serve as a guide for approximating likelihood ratios with neural network classifiers in the wider scientific community, and particularly within the domains of particle physics and astrophysics.

This paper is organized as follows. In Section \ref{sec:learning}, we summarize the theoretical foundation for learning likelihood ratios with neural network classifiers. In Section \ref{section:univariate_methods}, we present a series of studies focused on optimizing likelihood ratio estimation for one-dimensional Gaussian distributions where the true likelihood ratio is exactly known. In Section \ref{sec:multivariate}, we extend these studies to multi-dimensional Gaussian distributions. In Section \ref{sec:hep}, we present some more realistic examples using simulated high-energy physics data where the true likelihood ratio is approximated using a Normalizing Flow model \cite{normalizingflow}. Finally, we summarize our conclusions and recommendations for further studies in Section \ref{sec:conclusions}.

\section{Learning Likelihood Ratios}
\label{sec:learning}

Let the parameters \(\theta_0\) and \(\theta_1\) define two distributions, \(p(x\mid \theta_0)\) and \(p(x\mid \theta_1)\), as described in Section II.A. of \cite{Nachman_2021}. The goal is to determine or approximate the likelihood ratio
\begin{equation}
    \lr(x) = \frac{p(x\mid \theta_0)}{p(x\mid\theta_1)}
\end{equation}
between the two distributions.

Consider the general loss functional that depends on a learnable function \(f: \R^n \to \R\) and rescaling functions \(A: \R \to \R\) and \(B: \R \to \R\):

\begin{equation}
\begin{aligned} 
    L[f] &= -\int \d{x} \bigg(p(x \mid \theta_0) A(f(x))\\
    &\phantom{= -\int\d{x} \bigg(}+ p(x \mid \theta_1) B(f(x))\bigg)\,.
\end{aligned}
\end{equation}
We can take the functional derivative of the loss functional to show that the extremum can be transformed to obtain the likelihood ratio:
\begin{align}
    \frac{\delta L}{\delta f} &= -\frac{\partial}{\partial f}\Big( p(x\mid \theta_0) A(f(x)) + p(x\mid \theta_1) B(f(x))\Big)\\
    &\begin{aligned}
        &=- \bigg( p(x\mid\theta_0)A'(f(x)) \cdot f'(x)\\
        &\phantom{=- \bigg(} + p(x\mid \theta_0) B'(f(x)) \cdot f'(x) \bigg)
    \end{aligned}\\
    &= 0 \iff -\frac{B'(f(x))}{A'(f(x))} = \frac{p(x \mid \theta_0)}{p(x \mid \theta_1)} = \lr(x).
\end{align}
Given that \(-B'(f) / A'(f)\) is a monotonic rescaling of \(f\) and \(L[f]\) is convex, the learned function \(f\) is an optimal classifier.

In this paper, we first consider the four loss functionals defined by the rescaling functions in Table \ref{tab:losses}. While this is by no means an exhaustive list of all possible loss functionals, it includes a diverse array of different loss configurations. As detailed in Sec.~\ref{section:families}, we also consider generalized forms of two of these four loss functionals.
\begin{table}[ht]
\def\arraystretch{2} 
\centering
\begin{tabular}{m{48mm}cc}
Loss Name                     & $A(f)$                & $B(f)$      \\ \hhline{===}
Binary Cross-Entropy          & $\text{ln}(f)$             & $\text{ln}(1-f)$ \\
Mean Squared Error            & $-(1-f)^2$            & $-f^2$      \\
Maximum Likelihood Classifier & $\text{ln}(f)$             & $1-f$       \\
Square Root                   & $-\frac{1}{\sqrt{f}}$ & $-\sqrt{f}$
\end{tabular}
\captionof{table}{The rescaling functions \(A\) and \(B\) used to assemble the four different loss functionals considered.\label{tab:losses}}
\end{table}

A neural network parametrizes the learned function \(f\) as \(\phi(z)\), where \(z\) is the pre-activation output of the network and \(\phi\) is the final activation function. For the binary cross entropy (BCE) and mean squared error (MSE) losses,
\begin{equation} \label{eq: odds}
    \lr(x) = -\frac{B'(f)}{A'(f)} = \frac{f}{1 - f},
\end{equation}
so the likelihood ratio is the odds ratio of the learned function. That is, minimizing the BCE and MSE losses defines a classifier that computes
\begin{equation}
    \maxi_f L[f] = \frac{p(x \mid \theta_0)}{p(x \mid \theta_0) + p(x \mid \theta_1)} \in (0, 1).
\end{equation} 
To parametrize \(f\) such that the likelihood ratio is non-negative, we require that \({\phi: \R \to (0, 1)}\). 

However, for the maximum likelihood classifier (MLC) and square root (SQR) losses, 
\begin{equation} \label{eq: pure}
    \lr(x) = -\frac{B'(f)}{A'(f)} = f,
\end{equation}
so the likelihood ratio is the learned function, without transformation.
\begin{equation}
    \maxi_f L[f] = \frac{p(x \mid \theta_0)}{p(x \mid \theta_1)}
\end{equation}
In this case the loss-minimizing classifier computes the likelihood ratio \(\lr(x) \in (0, \infty)\). The requirement on \(\phi\) is that \(\phi: \R \to (0, \infty)\).

\section{Univariate Gaussians}
\label{section:univariate_methods}
In our first case study, we consider two Gaussian distributions with slightly different means and unit variances: $X_{0} \sim \text{Normal}(+0.1, 1)$ and $X_{1} \sim \text{Normal}(-0.1, 1)$. We also considered univariate Beta and Gamma distributions---these results can be found in Appendix \ref{sec:nongaussian}.

While one could in principle use Boosted Decision Trees (BDTs) instead of neural networks for the classifiers, we found that neural networks outperformed BDTs across a variety of test cases, as shown in Appendix \ref{sec:bdts}. All of our classifiers are therefore implemented as neural networks using \textsc{Keras} \cite{keras} with a \textsc{TensorFlow} \cite{tensorflow2015-whitepaper} backend and \textsc{Adam} \cite{adam} optimizer. Each classifier consists of three hidden layers with 64, 128, and 64 nodes, sequentially. Rectified Linear Unit (ReLU) activation functions are used for the intermediate layers, with the activation for the output layer depending on the loss used to train the neural network and the parametrization being tested. Each of the three hidden layers is followed by a dropout layer with dropout probability of 10\%. 

Unless otherwise stated, the networks were trained with 1,000,000 samples (750,000 used for training and 250,000 used for validation). 100,000 separate samples were used to evaluate the networks' performances (in particular, to calculate their mean absolute errors). Each network was trained for up to 100 epochs with a batch size of 10\%, as in \cite{Nachman_2021}. If the validation loss did not decrease for 10 consecutive epochs, the training was stopped (early stopping with a patience of 10). No detailed hyperparameter optimization was done.

\subsection{Na\"ive Implementation}
\subsubsection{Motivation}
The na\"ive parametrization for \(\phi(z)\) in the case of the BCE and MSE losses is \(\phi = \sigma\), the logistic function commonly used as the activation for classification tasks. In the case of the MLC and SQR losses, the most common parametrization would be \(\phi = \text{ReLU}\), the rectified linear unit activation. We chose these parametrizations for our na\"ive implementation.

To better understand how these common parametrizations of the classifiers affect their ability to learn the likelihood ratio, we implemented neural network architecture with each of the four losses, trained them to classify between the two Gaussian distributions. Since the true likelihood ratio is known, we can compare how well each of the four classifiers learns the likelihood ratio function.

\subsubsection{Methods}

We implemented each classifier using an identical neural network architecture, differing only in the final activation, which acted as either the logistic (for the BCE and MSE classifiers) or ReLU (for the MLC and SQR classifiers) parametrizations for the learned function.

We then trained each of the four classifier architecture on the dataset 100 times each, using the classifier's corresponding loss functional. Each classifier was evaluated on the interval \((-6, 6)\) and transformed into the likelihood ratio over that same interval using the appropriate transformation from equations \ref{eq: odds} and \ref{eq: pure}. We averaged the resulting 100 predictions for the likelihood ratio.

To numerically compare the performances of different classifiers in learning the likelihood ratio, we computed their empirical mean absolute errors over 100,000 samples. For \(\hat{\lr}\) the estimated likelihood ratio, the mean absolute error is defined as
\begin{equation}\label{eq:mae}
    \mae[\hat{\lr}] = \E\left[ \big\vert \lr(X) - \hat{\lr}(X) \big\vert \right].
\end{equation}
We computed this for each classifier as an empirical average over the 100 different likelihood ratio predictors to get a numerical measure of how well each predictor approximated the likelihood ratio.

Next, we examined how varying the amount of data upon which the classifiers were trained affected their performance. In particular, for each loss, we trained 100 classifiers for each \(N \in \{10^2, 10^3, 10^4, 10^5, 10^6, 10^7\}\). For each value of \(N\), \(0.75N\) observations were used for training and \(0.25N\) observations were used for validation. The value of \(N = 10^6\) corresponds to our default sample size. As before, 100,000 samples were used to estimate the MAE for each value of \(N\).

\subsubsection{Results}
\begin{figure}[ht]
    \centering
    \includegraphics{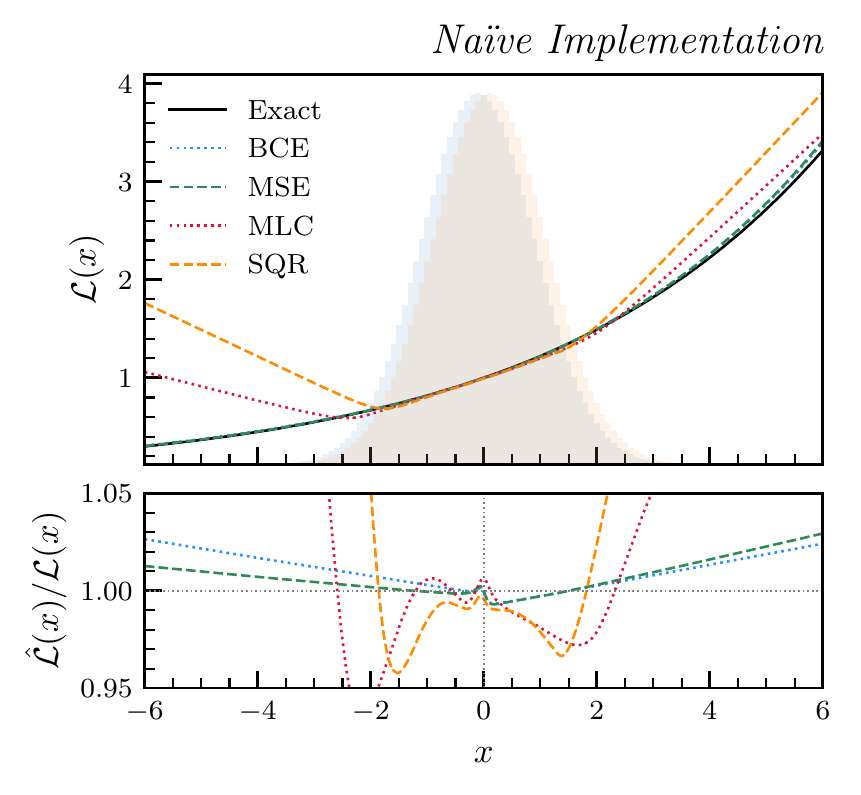}
    \caption{Average likelihood ratio fits for the four different losses. The MAEs are 0.0083, 0.0081, 0.0150, and 0.0254, for the BCE, MSE, MLC, and SQR likelihood ratio models, respectively.}
    \label{fig:naive_fit}
\end{figure}

Figure \ref{fig:naive_fit} displays the likelihood ratio fits averaged over 100 models for each of the four classifiers, compared against the true likelihood ratio. The largest deviations here are in regions far outside the bulk of the training data, where the models will largely be extrapolating. We are primarily concerned with evaluating the likelihood ratio approximation where the data has good coverage: approximately $x \in [-3,3]$. 

\begin{figure}[ht]
    \includegraphics{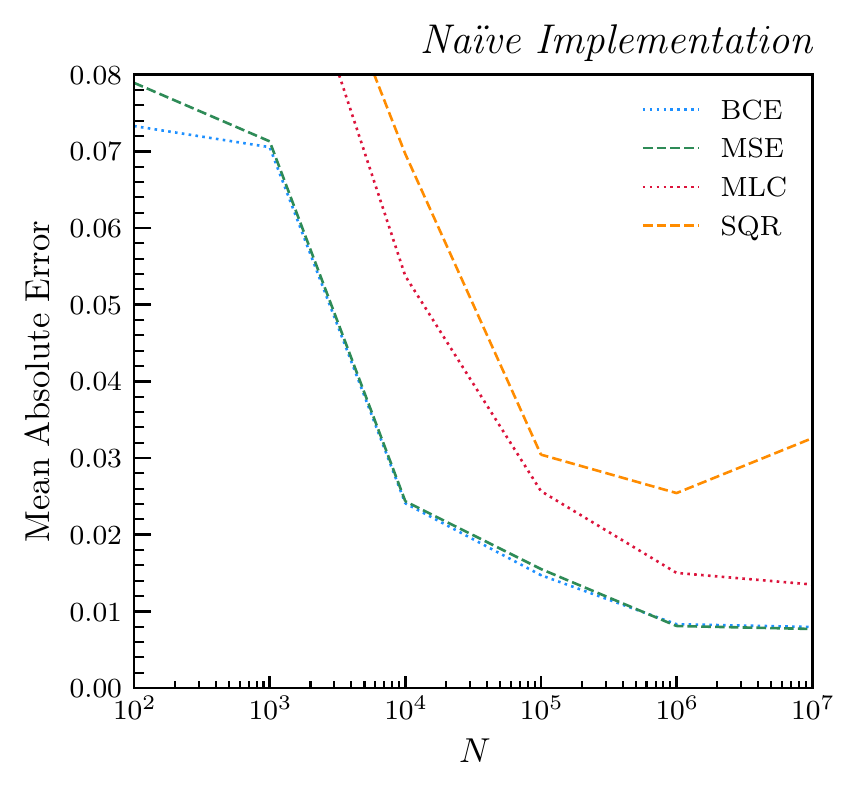}
    \caption{Mean absolute errors computed for the four different losses trained with increasingly larger sample sizes $N$.}
    \label{fig:naive_sizes}
\end{figure}

In Fig.~\ref{fig:naive_sizes}, we show how the expected error for classifiers trained with each choice of loss functional decreases as the sample size increases. 

\subsubsection{Discussion}
The four losses result in similarly performing fits near \(x = 0\); however, the MLC and SQR losses rapidly diverge from the true likelihood ratio in regions for which there is little data coverage. By comparison, the BCE and MSE perform much better, staying within 3\% of the true likelihood ratio even in regions far outside the bulk of the data ($|x| > 4$). 

The performance of these classifiers varies with the size of the training dataset $N$. For relatively small training sample sizes ($N<1000$), the scale of the mean absolute error is dominated by the inductive bias present in each activation function: BCE and MSE losses (both using $\sigma(z)$ activation) are nearly identical in the magnitude of their MAE, while MLC and SQR losses (both using $\text{ReLU}(z)$ activation) are similarly clustered. As $N$ increases, the MLC and SQR classifier performances approach those of the BCE and MSE classifiers. However, even for values of \(N\) larger than \(10^5\), the SQR classifier's MAE remains at least 0.015 above the average performance of the BCE/MSE classifiers. 

\subsection{\texorpdfstring{Parametrizing \(f\)}{Parametrizing f}}
\label{sec:univar_b}
The parametrization of the learned function can be adjusted. In the na\"ive implementation, the BCE and MSE neural networks use a logistic activation function, while the MLC and SQR neural networks use a ReLU activation function.

Let \(z(x)\) be the function that the neural network represents. Then \(f = \phi(z)\) is our classifier, where \(\phi\) is some parametrization of the learned function. In the cases described before, we have either \(f = \sigma(z)\) (for BCE and MSE) or \(f = \text{ReLU}(z)\) (for MLC and SQR). 

However, for BCE and MSE, any function \(\phi : \R \to (0, 1)\) will suffice. Two readily available such functions are the Gaussian CDF and arctangent, adjusted to the appropriate range:
\begin{align}
    f(z) &= \Phi(z) = \int_0^{z} \frac{1}{\sqrt{2\pi}} e^{-\frac{1}{2}x^2} \d{x},\\
    f(z) &= \frac{1}{\pi} \left(\arctan{z} + \frac{\pi}{2}\right).
\end{align}

\begin{figure*}[h]
    \centering
    \begin{subfigure}[b]{0.48\linewidth}
        \centering
        \includegraphics{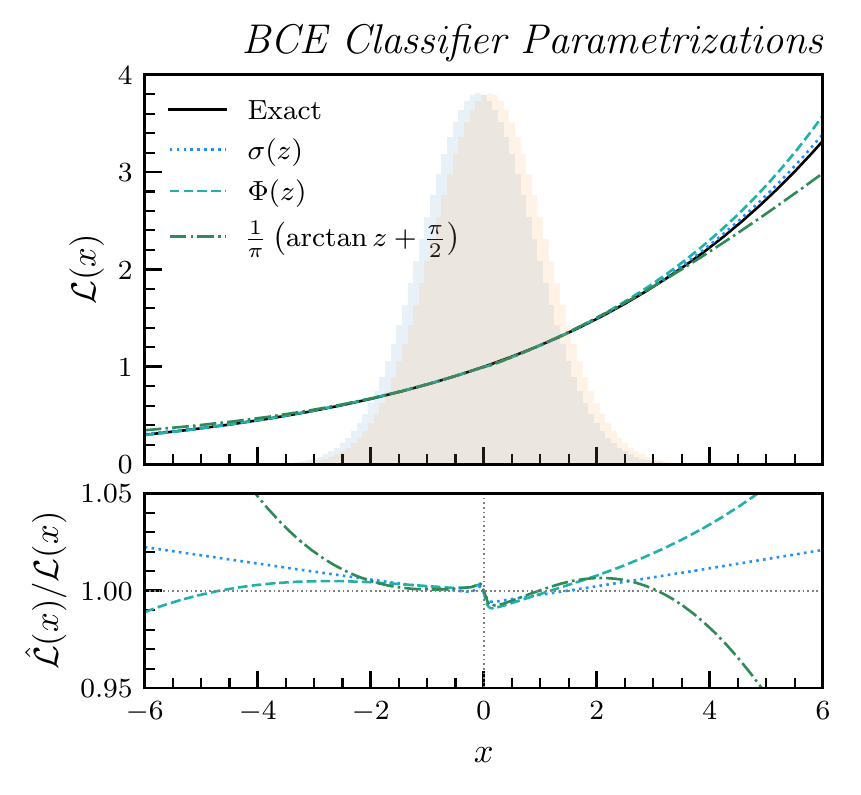}
        \caption{}
    \end{subfigure}
    \begin{subfigure}[b]{0.48\linewidth}
        \centering
        \includegraphics{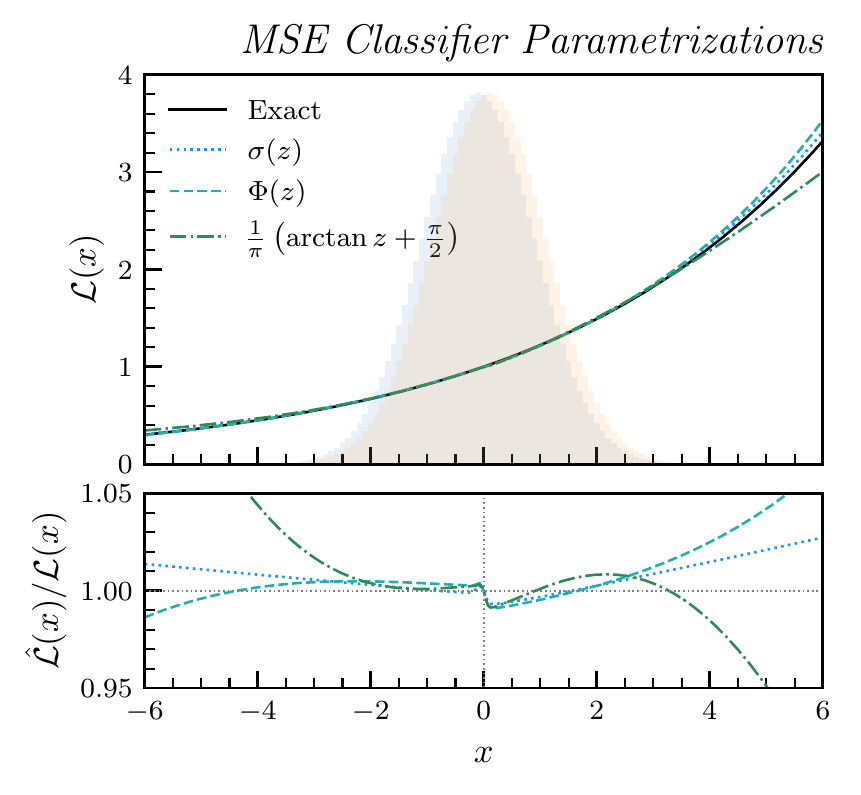}
        \caption{}
    \end{subfigure}
    \caption{Parametrizations of $f$ for the BCE and MSE losses. (a) The average likelihood ratio fits of the logistic, Gaussian CDF, and arctangent parametrizations for the BCE loss, with mean absolute errors 0.0081, 0.0113, and 0.0089, respectively. (b) The average likelihood ratio fits of the logistic, Gaussian CDF, and arctangent parametrizations for the MSE loss, with mean absolute errors 0.0081, 0.0110, and 0.0010, respectively.}
    \label{fig:bce_mse_c}
\end{figure*}

\begin{figure*}[h]
    \begin{subfigure}[b]{0.48\linewidth}
        \includegraphics{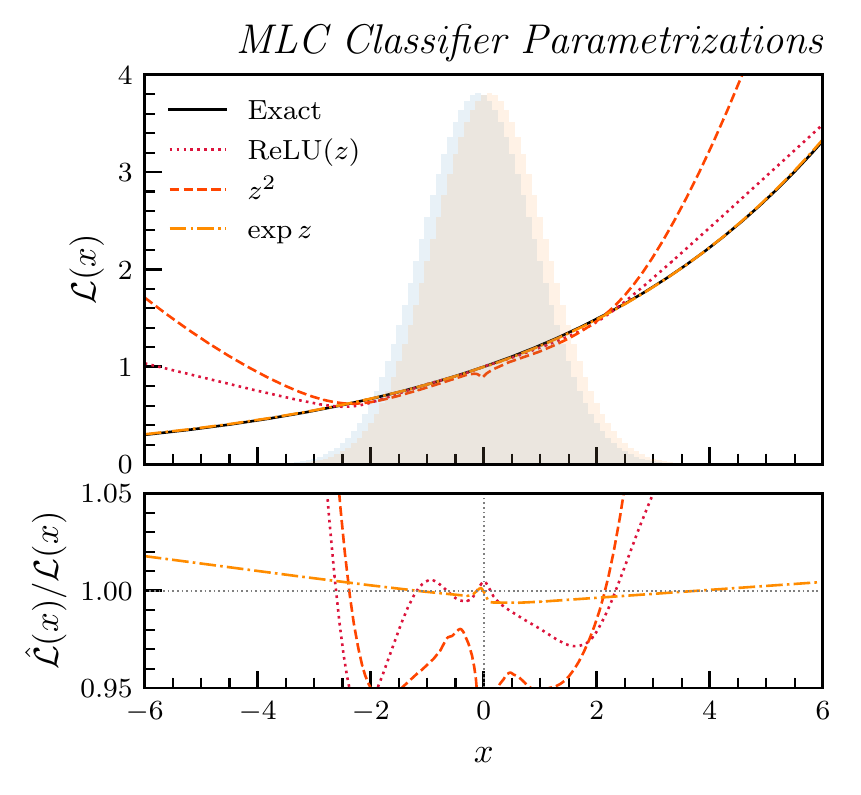}
        \caption{}
    \end{subfigure}
    \begin{subfigure}[b]{0.48\linewidth}
        \includegraphics{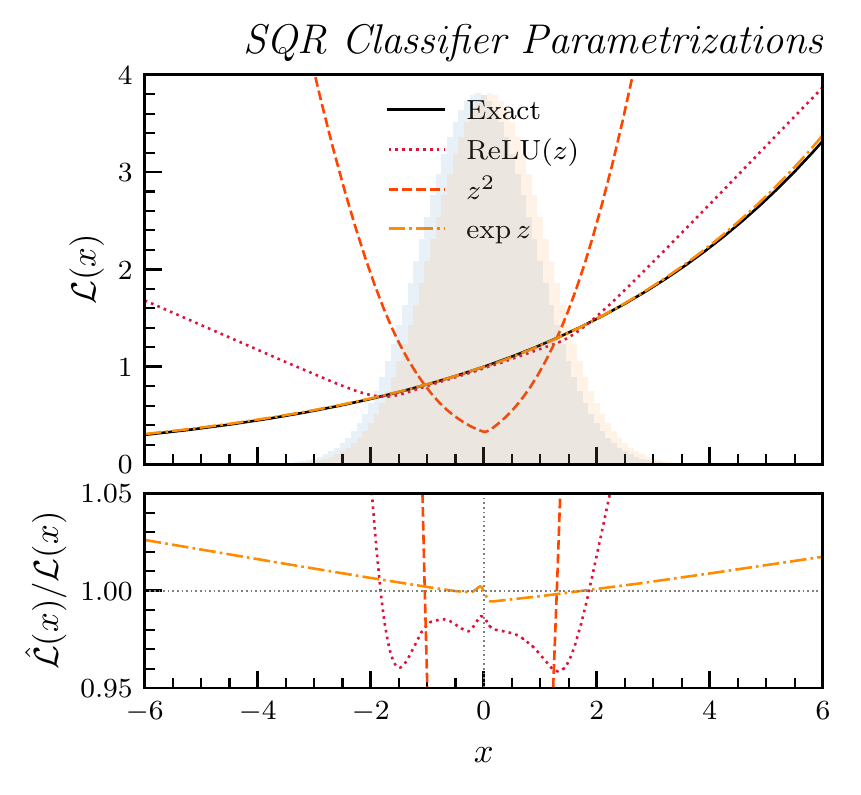}
        \caption{}
    \end{subfigure}
    \caption{Parametrizations of $f$ for the MLC and SQR losses. (a) The average likelihood ratio fits of the ReLU, square, and exponential parametrizations for the MLC loss, with mean absolute errors 0.0148, 0.0684, and 0.0083, respectively. (b) The average likelihood ratio fits of the ReLU, square, and exponential parametrizations for the SQR loss, with mean absolute errors 0.0367, 0.6756, and 0.0075, respectively.}
    \label{fig:mlc_sqr}
\end{figure*}

In Fig.~\ref{fig:bce_mse_c}, we show the likelihood ratio fits, averaged over 100 models, for the logistic, Gaussian CDF, and arctangent parametrizations of the BCE and MSE classifiers. In both cases, the default logistic parametrization performs the best, followed closely by the Gaussian CDF and arctangent parametrizations. This result is not surprising, as the logistic function is known to be well-suited for classification.

For the MLC and SQR losses, we instead require any function \(\phi: \R \to (0, \infty)\). While the ReLU function is the default, there are other functions with such ranges, including:
\begin{align}
    f(z) &= z^2,\\
    f(z) &= \exp{z}.
\end{align}

Figure \ref{fig:mlc_sqr} displays the results of comparing the performances of the MLC and SQR losses in training classifiers with these parametrizations. The performances of the three parametrizations between the two losses are the same: in this case, the exponential parametrization performs remarkably better than the ReLU parametrization, and square parametrization performs the worst amongst all three. It is worth noting that exponentially parametrizing the neural network is identical to training a neural network to learn the log likelihood ratio.

\subsection{Generalized Loss Families}
\label{section:families}

\subsubsection{Motivation}
The MSE and SQR loss functionals are easily generalizable to a parametric family of loss functionals. While there are several possible parametrizations\footnote{For example, to enforce non-singular behavior at $r=0$ for SQR, one could consider $A(f) = (1 - f^{-\frac{r}{2}})/|r|$ and $B(f) = (1 - f^{\frac{r}{2}})/|r|$. Another interesting parametrization is $A(f) = (f^q - 1)/q$ and $B(f) = 1-f^{(q+1)}/(q+1)$, which is minimized at $q=1$.} to choose from, we select the following for simplicity: for the MSE loss, we consider a power parameter \(p \in \R\), where $p=2$ is the default value, and for the SQR loss, we consider a root parameter \(r \in \R\), where $r=1$ is the default value. This yields the two families of losses presented in Table \ref{tab:families}.

Since the rescaling functions \(A\) and \(B\) have changed, the likelihood ratio recovered from \(f\) changes as well. 


\begin{figure*}[ht]
    \centering
    \begin{subfigure}[b]{0.48\linewidth}
        \centering
        \includegraphics{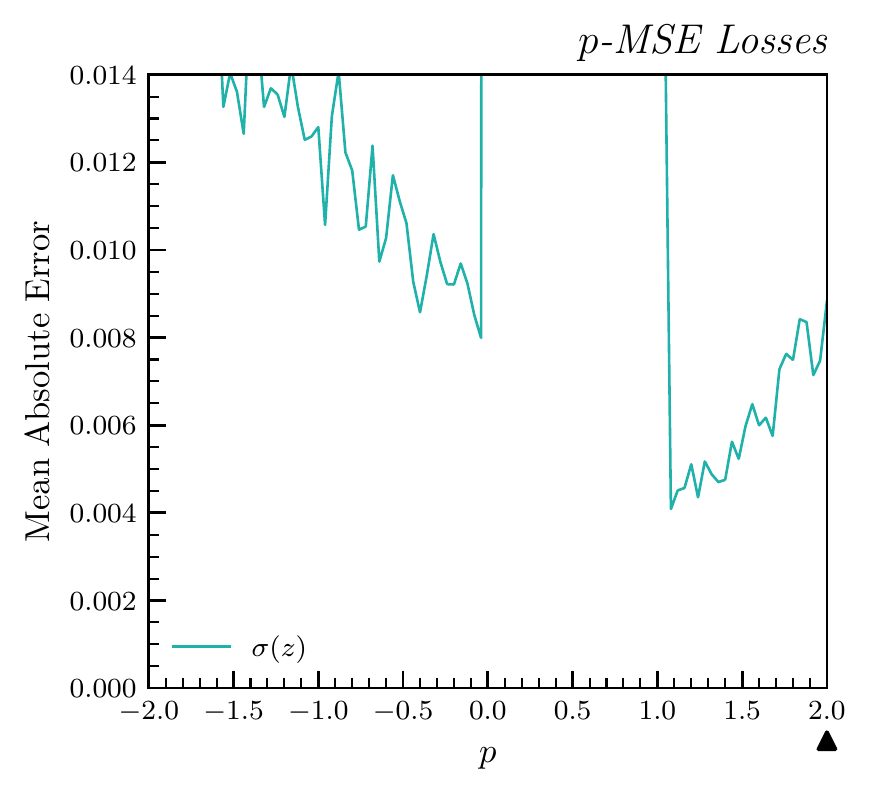}
        \caption{}
    \end{subfigure}
    \begin{subfigure}[b]{0.48\linewidth}
        \centering
        \includegraphics{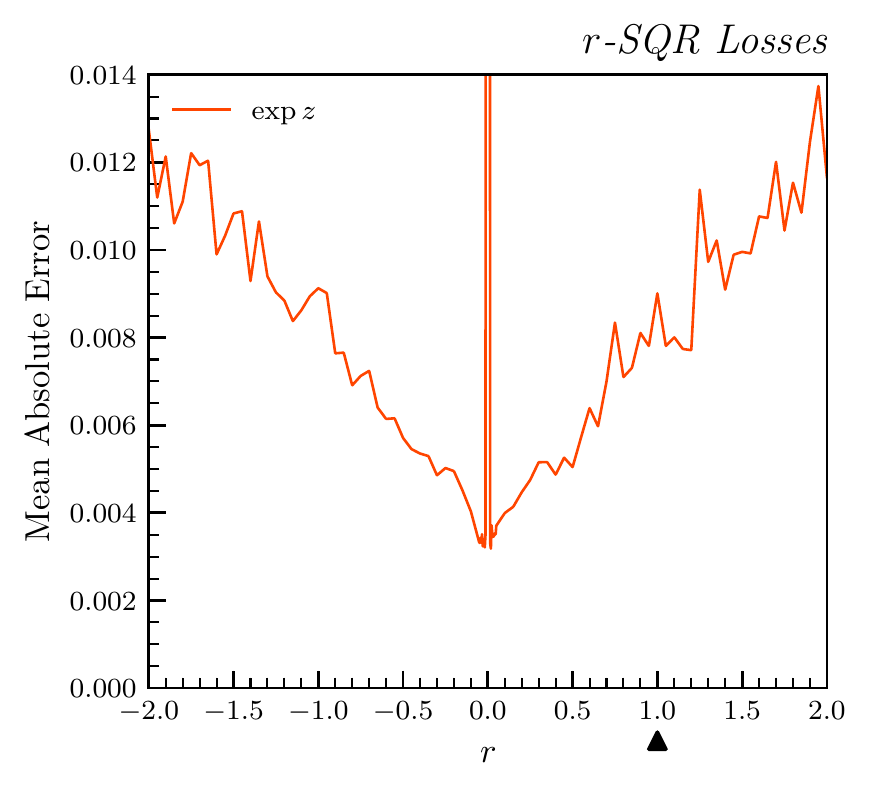}
        \caption{}
    \end{subfigure}
    \caption{(a) The mean absolute errors averaged over models trained on the generalized MSE loss family for the logistic parametrization. The mean absolute error is minimized at \(p^* = 1.08\), but we choose the second-lowest value (\(p^* = 1.24\)) for stability, i.e. avoiding the steep increase in MAE near \(p = 1\). The arrow indicates the typical choice of $p=2$ for MSE loss. (b) The mean absolute errors averaged over models trained on the generalized SQR loss family for the exponential parametrization. The mean absolute error was smallest at \(r^* = 0.018\). The arrow indicates the typical choice of $r=1$ for SQR loss.}
    \label{fig:mse_sqr_ab}
\end{figure*}

\begin{table}[ht]
\def\arraystretch{2} 
\centering
\begin{tabular}{rcc}
Loss Name                     & $A(f)$                & $B(f)$      \\ \hhline{===}
\(p\)-MSE            & $-(1-f)^p$            & $-f^p$      \\
\(r\)-SQR                   & $-f^{-\frac{r}{2}}$   & $-f^{\frac{r}{2}}$
\end{tabular}
\captionof{table}{The generalization of the MSE and SQR loss functionals to entire families of losses. Values of \(p = 2\) and \(r = 1\) correspond to the original definitions of the loss functionals.\label{tab:families}}
\end{table}

For the \(p\)-MSE losses, for \(p \notin (0, 1)\),

\begin{align}
    \lr(x) &= -\frac{B'(f)}{A'(f)} =-\frac{-pf^{p-1}\cdot f'}{p(1-f)^{p-1} \cdot f'}\\ 
    &= \left(\frac{f}{1 - f}\right)^{p-1}.
\end{align}

We exclude the case where \(p \in (0, 1)\) since the corresponding loss functional is not convex, and as such the likelihood ratio trick no longer works.

And for the \(r\)-SQR losses, for \(r \neq 0\),

\begin{align}
    \lr(x) &= -\frac{B'(f)}{A'(f)} = -\frac{-\frac{r}{2} f^{\frac{r}{2} - 1}\cdot f'}{\frac{r}{2} f^{-\frac{r}{2} - 1}\cdot f'}\\
    &=f^r.
\end{align}

The case where \(r = 0\) is excluded since the corresponding loss functional is not strictly convex.

A whole family of losses arises from both of the two original losses, each loss still maintaining the property that the function that minimizes the corresponding functional can recover the likelihood ratio. In addition to comparing how the four original losses performed against one another, we can compare among the losses in each of these two loss families.

\subsubsection{Methods}
Since we were working over an uncountably infinite set of loss functionals, we decided to constrain our investigation to just the interval $[-2, 2]$. We scanned along the interval $[-2, 2]$; for each value $p$ which we looked at, we trained 20 logistically-parametrized models on the $p$-MSE loss functional corresponding to that value of $p$. Then we averaged the mean absolute errors of the 20 models together. 

We did the same for values of $r$ in the interval $[-2, 2]$ as well; in that case, the models were parametrized with the exponential activation function instead.

We expect that near $p^*=1$ and $r^*=0$, where the generalized loss functionals will resemble the MAE loss, the figure-of-merit of MAE will likely be minimized, too. Due to this intrinsic relationship between the choice of loss functional and figure-of-merit, we also considered two additional figures-of-merit for evaluating these scans: the Mean Ratio and the Null Statistic, defined as: 
\begin{equation}
    \text{Mean Ratio}[\hat{\mathcal{L}}] =  E\left[\hat{\mathcal{L}}(X) / \mathcal{L}(X)\right]
\end{equation}
\begin{equation}
     \text{Null Statistic}[\hat{\mathcal{L}}] = \Big|E_0[\mathcal{L}(X)] - E_0[\hat{\mathcal{L}}(X)]\Big|
\end{equation}
We found that the overall trends reported here using MAE were similar across these alternative figures-of-merit, though the trends were less dramatic when we used the Mean Ratio figure-of-merit.

\subsubsection{Results}
In Figure \ref{fig:mse_sqr_ab}, we show the performance of the classifiers trained by these losses when modifying their power and root parameters. The values $p^*$ and $r^*$ minimizing the MAE were $p^* = 1.08, 1.24$ (with $p^* = 1.24$ having a similar performance to that of $p^* = 1.08$ while being more numerically stable) and $r^* = 0.018$. 

\subsubsection{Discussion}
In \ref{fig:mse_sqr_ab}(a), we observe vertical features for \(p \in (0, 1)\). This is to be expected, as the likelihood ratio trick does not apply in the range where the corresponding loss functional is non-convex. Similarly, the vertical feature in \ref{fig:mse_sqr_ab}(b) is due to the fact that for \(r = 0\), our loss functional is constant (\(L[f] = 1\)), and thus it is not strictly convex; therefore the likelihood ratio again does not work.

Values of \(p\) slightly less than \(0\) or slightly greater than \(1\) resulted in the smallest mean absolute errors, while values of \(r\) close to \(0\) resulted in the smallest mean absolute errors. 

This result was further investigated in Section \ref{section:simple} in a simple, two-dimensional classifier model.

\subsection{Optimized Implementation}
Altering the parametrization of the learned function \(f\) or using a more generalized loss functional yielded considerable increases in performance from the initial parametrizations and loss functionals. 

\begin{figure}[ht]
    \centering
    \includegraphics{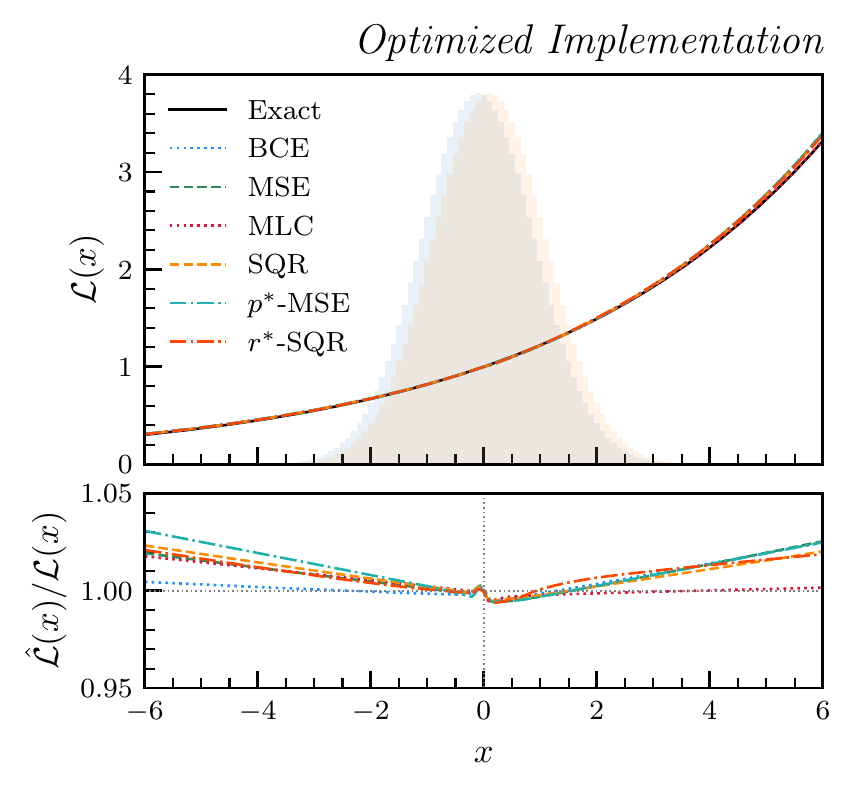}
    \caption{Average likelihood ratio fits for the different loss categories. The MAEs are 0.0079, 0.0045, 0.0077, 0.0034, 0.0046, and 0.0034, for the BCE, MSE, MLC, SQR, $p^*$-MSE, and $r^*$-SQR likelihood ratio models, respectively.}
    \label{fig:best_fit}
\end{figure}

\begin{figure}[ht]
    \centering
    \includegraphics{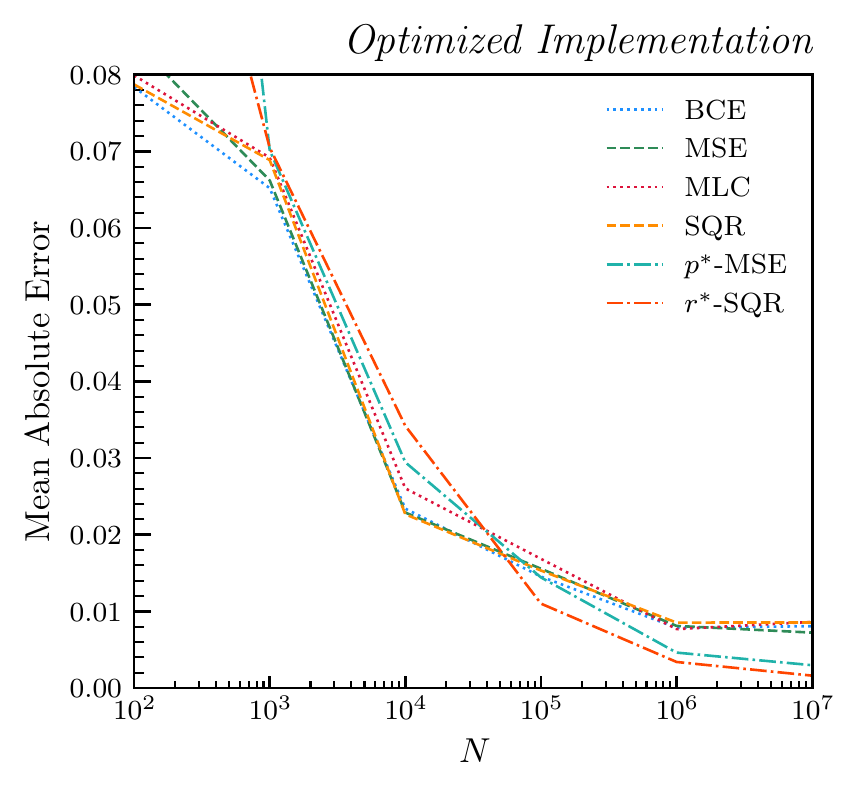}
    \caption{Mean absolute errors computed for the different loss categories trained with increasingly larger samples.}
    \label{fig:best_sizes}
\end{figure}

In Figs.~\ref{fig:best_fit} and \ref{fig:best_sizes}, we chose the best-performing parametrization for each loss (logistic for BCE and MSE; exponential for MLC and SQR), and, for the MSE and SQR, chose the best-performing loss functional from each loss family (\(p^* = 1.24\) for MSE and \(r^* = 0.018\) for SQR), and trained classifiers with each ``optimized" parametrization and loss. This was done 100 times for each parametrization/loss, and the resulting likelihood ratio models were averaged. 

In the na\"ive implementation, the BCE and MSE models performed the best, while the SQR model had an average error at least 0.015 larger than the other losses, even for large $N$. In the optimized implementation with $N=10^6$, all four loss functionals perform approximately the same, as shown in Figure \ref{fig:best_fit}. Figure \ref{fig:best_sizes} shows that for $N > 10^5$, the four optimized loss functionals continue to perform approximately equally well, but the new loss functionals $p^*$-MSE and $r^*$-SQR perform significantly better, reaching mean absolute errors about 2 to 4 times smaller than the other losses. The strong influence of the inductive bias of the activation function is also mitigated in the optimized implementation, as the losses are no longer grouped by activation function. 


\subsection{Simple Classifiers}
\label{section:simple}
\subsubsection{Motivation}
To better understand the behavior of the generalized loss models in Sec.~\ref{section:families}, we examined a much simpler classifier than the multi-layer fully-connected network used to train the models in this paper. This allows us to visualize the dynamics of each model, using numerical integration to compute the loss.  The model is 
\begin{equation}
    f(x) = \phi(ax + b),
\end{equation}
where \(a\) and \(b\) are the two weights of the model and \(\phi\) is its activation.

\begin{figure*}[ht]
    \centering
    \hspace*{.584cm}\includegraphics{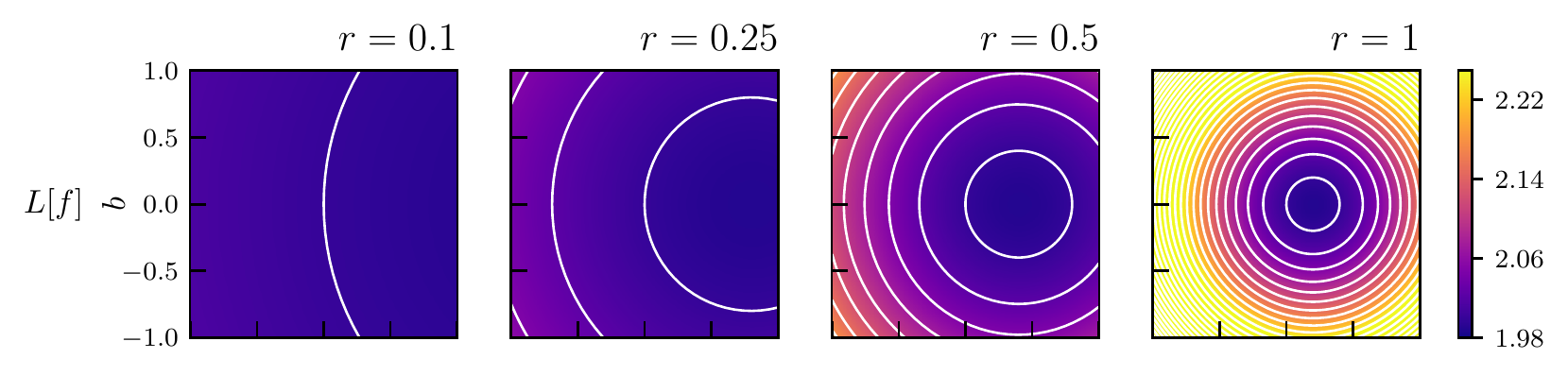}\\\vspace*{-.25cm}\includegraphics{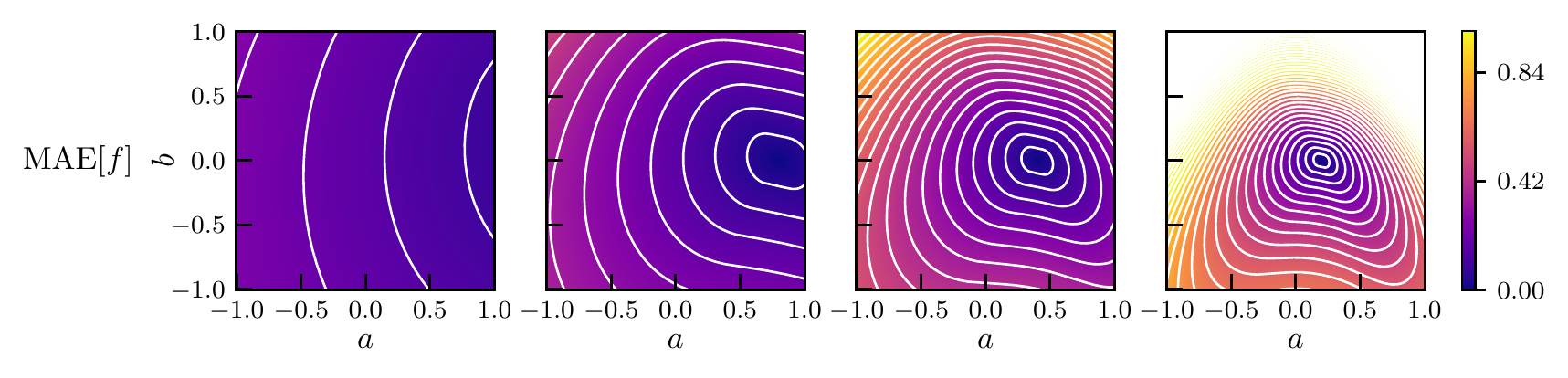}
\caption{Contour plots of the losses and MAEs of the two-dimensional SQR classifier \(f(x)=\exp{(ax + b)}\) over \([-1, 1]^2\). The first row plots the value of the loss functional \(L[f]\), obtained through numerical integration, on a grid of \((a, b)\) pairs over \([-1, 1]^2\) for various values of \(r\), with contours curves at increments of \(0.02\). The second row plots an empirically computed absolute error, \(\mae[f]\), over the same grid of points, for the same values of \(r\), with contour curves at increments of \(0.05\).}
\label{fig:landscapes}
\end{figure*}

In the case of the \(p\)-MSE model with the logistic parametrization, 
\begin{align}
    f_{\rm MSE}(x) &= \sigma(ax + b)\\
    \hat{\lr}_{\rm MSE}(x) &= \left(\frac{\sigma(ax + b)}{1 - \sigma(ax + b)}\right)^{p-1} = \left(e^{ax + b}\right)^{p-1}.
\end{align}
The exponentially-parametrized \(r\)-SQR model is
\begin{align*}
    f_{\rm SQR}(x) &= e^{ax + b}\\
    \hat{\lr}_{\rm SQR}(x) &= \left(e^{ax+b}\right)^r.
\end{align*}
As a result, only the analysis of one of the two models is necessary, since the resulting likelihood ratio model is the same for \(r = p - 1\). In particular, we will analyze the \(r\)-SQR model, keeping in mind that for the model MAEs, the results will be identical for \(p = r + 1\).

We continue working with \(X_0 \sim \text{Normal}(+0.1, 1)\) and \(X_1 \sim \text{Normal}(-0.1, 1)\). The exact likelihood ratio is given by
\begin{equation}
    \lr(x) = e^{0.2x},
\end{equation}
so the two-dimensional classifier will yield an exact solution at 
\begin{equation}
a^* = \frac{0.2}{r}, \qquad b^* = 0.
\end{equation}

\subsubsection{Methods}
To better understand how the parameter \(r\) affects the optimization landscape, we first created a grid with fineness \(0.005\) of \((a, b)\) pairs in the box \([-1, 1]^2\):
\begin{equation}
    B = \frac{1}{200}\Z^2 \cap [-1, 1]^2.
\end{equation}
The loss functional for a particular value of \(r\), \(L_r\) is given by
\begin{equation} 
    L_r[f] = \int \d{x} \bigg(p(x \mid \theta_A) f(x)^{-\frac{r}{2}}+ p(x \mid \theta_B) f(x)^{\frac{r}{2}}\bigg)
\end{equation}
Then, we visualized the loss landscape as the contour plot of \(L_r\) over the set of classifiers \(F = \{(e^{ax +b})^r: (a, b) \in [-1, 1]^2\}\) for different values of \(r\). The loss functional \(L_r\) was computed via numerical integration.

\subsubsection{Results}
Figure \ref{fig:landscapes} displays in the first row the resulting contour plots for \(r \in \{0.1, 0.25, 0.5, 1\}\). Drawn over each plot, in white, are level sets of the loss at increments of 0.02. 

\subsubsection{Discussion}

While the actual values of the losses are not comparable between different values of \(r\), since each value of \(r\) corresponds to a different loss functional, it is clear that the loss functional becomes increasingly steep as \(r\) increases. As expected, as \(r \to \infty\), \(a^* \to 0\), and as \(r \to 0\), \(a^* \to \infty\). In particular, the loss landscape of \(r = 0.1\) is shaped like an extremely shallow pool, indicating that there is a large space of classifiers with close to optimal performance. The minimum value \(a^* = 2\) is not visible in the box, since small values of \(r\) correspond to large values of \(a^*\). On the other hand, the loss landscape of \(r = 1\) is much steeper, with a minimum at \(a^* = 0.2\) around which the landscape quickly increases to high loss values.

However, since loss values between values of \(r\) are incomparable, it is unclear how the loss reflects the actual performance of the likelihood ratio model. In particular, given two classifiers \(f\) and \(g\) with \(L_{r}[f] < L_{s}[g]\), \(r \neq s\), we cannot be sure that \(f\) will yield a better likelihood ratio model than \(g\), since \(L_{r}\) and \(L_{s}\) are different loss functionals.

To this end, we visualized the error landscape as the contour plot of \(\mae\) over the same set of classifiers \(F = \{(e^{ax + b})^r: (a, b) \in [-1,1]^2\}\). Since the MAE is computed only from the expected absolute difference between a predicted likelihood ratio \(\hat{\lr}\) and the true likelihood ratio \(\lr\), we can compare across different values of \(r\) to see which values of \(r\) result in easily obtainable well-performing classifiers.

The second row of Figure \ref{fig:landscapes} displays these error contour plots; indicated in white are the level sets of the error at increments of 0.05. For each of these, we can see that the error is zero at \((a^*, 0)\) and increases radially outwards from the minimum. The shape of the loss landscapes reflect the true nature of the performance of the classifiers; for \(r = 0.1\), we still have a shallow pool of many well-performing classifiers, whereas for \(r = 1\), there is a small set of well-performing classifiers around which the classifiers begin to perform much worse. That is to say, for small values of \(r\), there are many classifiers that perform well at modeling the likelihood ratio. It may be harder to find the true minimum, but most classifiers have comparable performance. On the other hand, for large values of \(r\), the loss landscape is steep with few classifiers with decent performance. Slight perturbations around the minimum correspond to large errors.

\section{Multivariate Gaussians}
\label{sec:multivariate}
\subsection{\texorpdfstring{Parametrizing \(f\)}{Parametrizing f}}
\subsubsection{Motivation}
A natural extension from the univariate Gaussians analysis in the previous section would be to multivariate Gaussians, wherein the setting is complicated by the higher dimensions, but we still have knowledge of the true likelihood ratio. To this end, we first established five different case studies of different Gaussian arrangements to examine in our multivariate analysis.

The first case study, labeled ``Vertical," corresponds to independent Gaussians with variance 1, and means at a distance of \(0.2\), as in the univariate case.

In this case, the background distribution is more likely over the right half-plane, whereas the signal distribution is more likely over the left half-plane.

\begin{align}
    X_0 &\sim \text{Normal}\left( \begin{bmatrix} +0.1 \\ 0 \end{bmatrix}, \begin{bmatrix} 1 & 0 \\ 0 & 1 \end{bmatrix} \right)\\
    X_1 &\sim \text{Normal}\left( \begin{bmatrix} -0.1 \\ 0 \end{bmatrix}, \begin{bmatrix} 1 & 0 \\ 0 & 1 \end{bmatrix} \right)
\end{align}

The next case study, ``Slant," simply rotates the vertical case study by \(45^\circ\). This results in the same likelihood ratio as the vertical case, except rotated by \(45^\circ\).

\begin{align}
    X_0 &\sim \text{Normal}\left( \begingroup \renewcommand*{\arraystretch}{1.75} \begin{bmatrix} +\frac{0.1}{\sqrt{2}} \\ -\frac{0.1}{\sqrt{2}} \end{bmatrix}\endgroup, \begin{bmatrix} 1 & 0 \\ 0 & 1 \end{bmatrix}\right)\\
    X_1 &\sim \text{Normal}\left( \begingroup \renewcommand*{\arraystretch}{1.75}\begin{bmatrix} -\frac{0.1}{\sqrt{2}} \\ +\frac{0.1}{\sqrt{2}} \end{bmatrix}\endgroup , \begin{bmatrix} 1 & 0 \\ 0 & 1 \end{bmatrix}\right)
\end{align}

\begin{figure}[ht]
    \centering
    \includegraphics{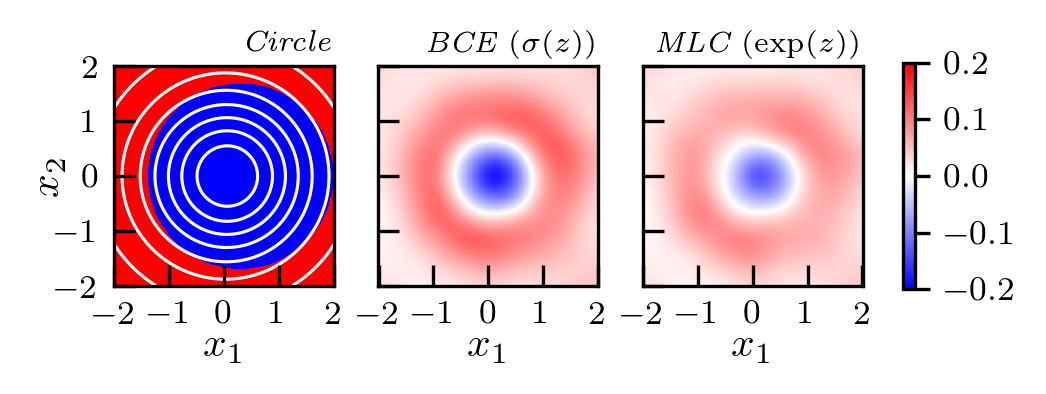}\\\vspace*{-.25cm}\includegraphics{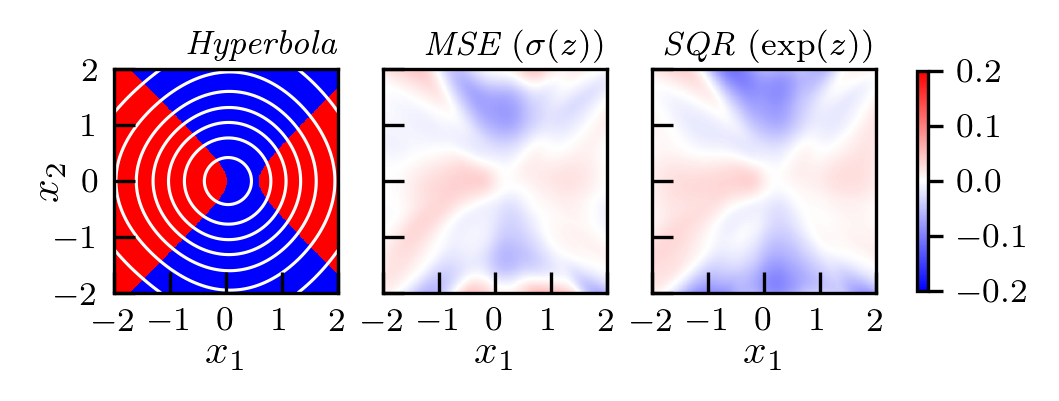}
    \caption{Two of the five multivariate Gaussian cases we examined, as well as some of the likelihood ratio model fits. The first row corresponds to the circle case, while the second row corresponds to the hyperbola case. The first column plots the likelihood structure of each case; red regions are regions where \(\lr(x) \le 1\), and blue regions are regions where \(\lr(x) > 1\). The second and third columns display contour plots of the mean absolute error for some models trained with the various losses to learn the likelihood ratios. The plot is suggestively colored to show how the structure of the data corresponds to the structure in the likelihood ratio models.}
    \label{fig:mvn_cases}
\end{figure}

In ``Circle," we consider the case where the background distribution has low variance in comparison to the signal distribution. As a result, values close to the origin are more likely to be from the background, whereas values far from the origin are more likely to be from the signal. This likelihood structure is visualized in Figure \ref{fig:mvn_cases}.

\begin{align}
    X_0 &\sim \text{Normal}\left( \begin{bmatrix} +0.1 \\ 0 \end{bmatrix}, \begin{bmatrix} 1 & 0 \\ 0 & 1 \end{bmatrix} \right)\\
    X_1 &\sim \text{Normal}\left( \begin{bmatrix} -0.1 \\ 0 \end{bmatrix}, \begin{bmatrix} 2 & 0 \\ 0 & 2 \end{bmatrix} \right)
\end{align}

The ``Hyperbola" case study looks at the case when both the background and the signal have different variances in each coordinate. This results in a hyperbola-like likelihood structure, as visualized in Figure \ref{fig:mvn_cases}.

\begin{align}
    X_0 &\sim \text{Normal}\left( \begin{bmatrix} +0.1 \\ 0 \end{bmatrix}, \begin{bmatrix} 1 & 0 \\ 0 & 2 \end{bmatrix} \right)\\
    X_1 &\sim \text{Normal}\left( \begin{bmatrix} -0.1 \\ 0 \end{bmatrix}, \begin{bmatrix} 2 & 0 \\ 0 & 1 \end{bmatrix} \right)
\end{align}

Finally, ``Checker" looks at the case where the coordinates for the background are correlated and the coordinates for the signal are correlated.

\begin{align}
    X_0 &\sim \text{Normal}\left( \begin{bmatrix} +0.1 \\ 0 \end{bmatrix}, \begin{bmatrix} 2 & -\frac{1}{4} \\ -\frac{1}{4} & 1 \end{bmatrix} \right)\\
    X_1 &\sim \text{Normal}\left( \begin{bmatrix} -0.1 \\ 0 \end{bmatrix}, \begin{bmatrix} 2 & +\frac{1}{4} \\ +\frac{1}{4} & 1 \end{bmatrix} \right)
\end{align}

\subsubsection{Methods}
The methodology was similar to that done in Sec.~\ref{section:univariate_methods}. For each case study, we implemented all four classifiers with each of the three parametrizations. Each resulting classifier architecture was trained 100 times to minimize the corresponding loss functional. We evaluated each classifier on the box \([-2, 2]^2\), and we averaged the resulting 100 predictions for the likelihood ratio over that box. We used the MAE as the performance metric, again as an empirical average over 100,000 samples.

\subsubsection{Results}
\begin{figure*}[ht]
    \centering
    \includegraphics[width=\textwidth]{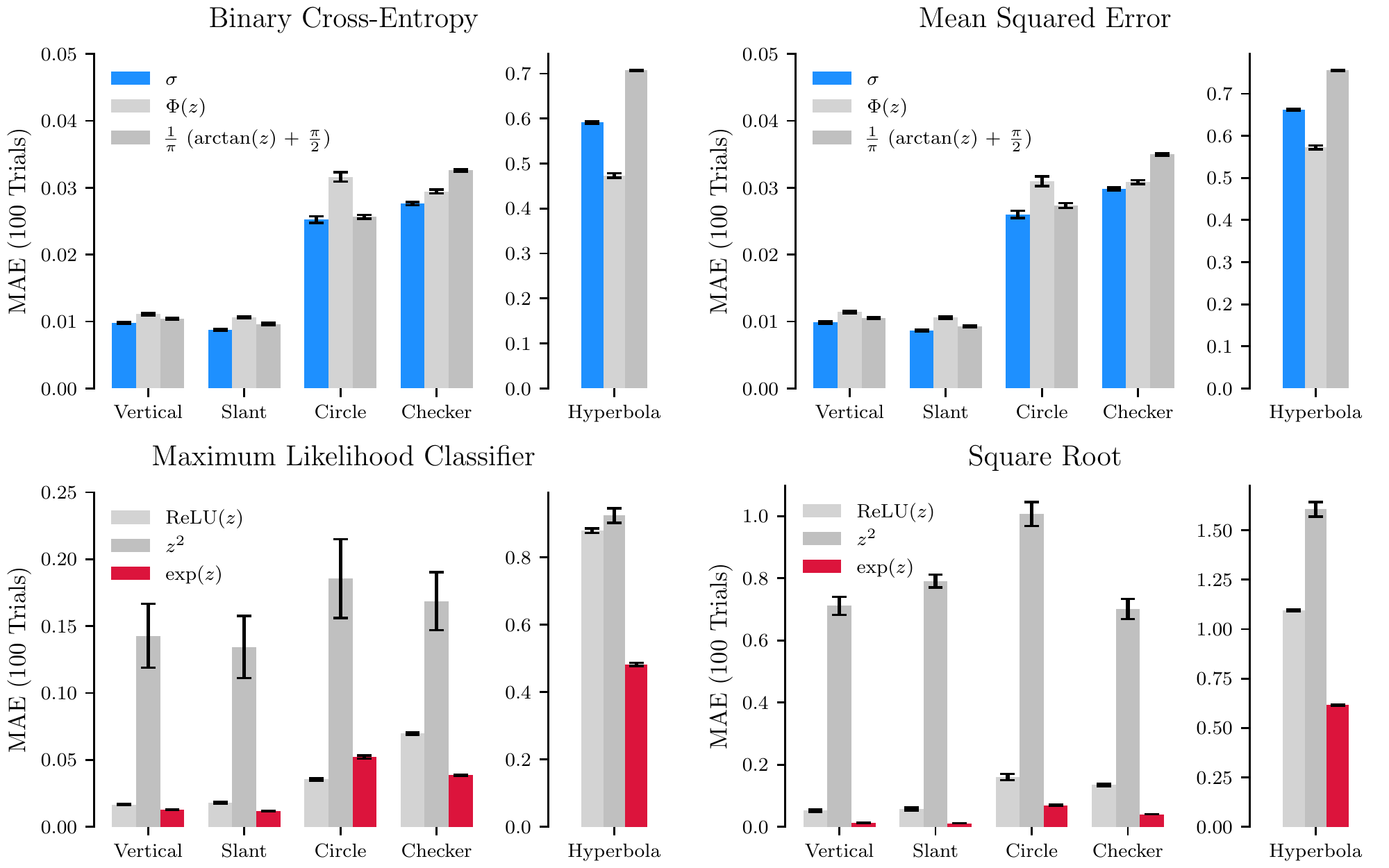}
    \caption{Mean absolute errors are compared for the four different losses considered (binary cross-entropy, mean squared error, maximum likelihood classifier, and square root), each with 3 different activation functions. For each loss, five different multivariate normal cases are studied: ``Vertical'', ``Slant'', ``Circle'', ``Checker'', and ``Hyperbola''. For each case study, the best performing parametrization for each loss is shown in either red or blue. Errors represent the standard deviation across 100 independent model trainings.}
    \label{fig:mvn_losses}
\end{figure*}

The resulting MAEs are shown in Figure \ref{fig:mvn_losses}. Some contour plots of the mean absolute errors of some of the different parametrizations are presented in Figure \ref{fig:mvn_cases}; the remaining contour plots are provided in Appendix \ref{sec:mvn_appendix}.

\subsubsection{Discussion}
In the univariate case, we found that the logistic and exponential parametrizations were uniformly the best parametrizations for the BCE/MSE and MLC/SQR losses, respectively. This trend is largely consistent with the results for these higher-dimensional cases. In all but one of the cases (Hyperbola), the logistic parametrization performed the best for the BCE/MSE losses. The exponential parametrization performed the best for all the cases for the MLC/SQR losses.

Unlike in Section \ref{section:univariate_methods}, once the optimal parametrizations are chosen for each of the four loss functionals, some differences persist in the performance of each loss. Across all five cases, the SQR loss yields the largest errors. For the Vertical and Slant cases, all four optimized loss functionals perform equally well, overlapping within one standard deviation. For the remaining cases (Checker, Circle, and Hyperbola), the optimized MLC loss with exponential parametrization performs significantly better than the other three optimized losses. 

It is striking to note that the MLC loss with exponential parametrization emerges as the best-performing loss configuration in some of the more complex datasets considered for these studies. The typical choice for a neural network classifier loss is arguably BCE. For the purposes of the likelihood ratio trick, however, we are interested in reinterpreting the classifier output to approximate the likelihood ratio, so it is possible that optimizing for raw classification performance alone is misguided. The MLC loss has the advantage of explicitly relating the signal and background probability distributions; in particular, the MLC loss can be intuitively understood to maximize the likelihood of \(\hat{\lr}(x)\) with respect to \(p(x \mid \theta_0)\) subject to the constraint that \(\hat{\lr}(x) p(x \mid \theta_1)\) is a probability distribution \cite{Nachman_2021}. Therefore, it may be a more natural choice for this particular application than the default BCE loss. 

\begin{figure*}[h]
    \centering
    \includegraphics[width=0.9\textwidth]{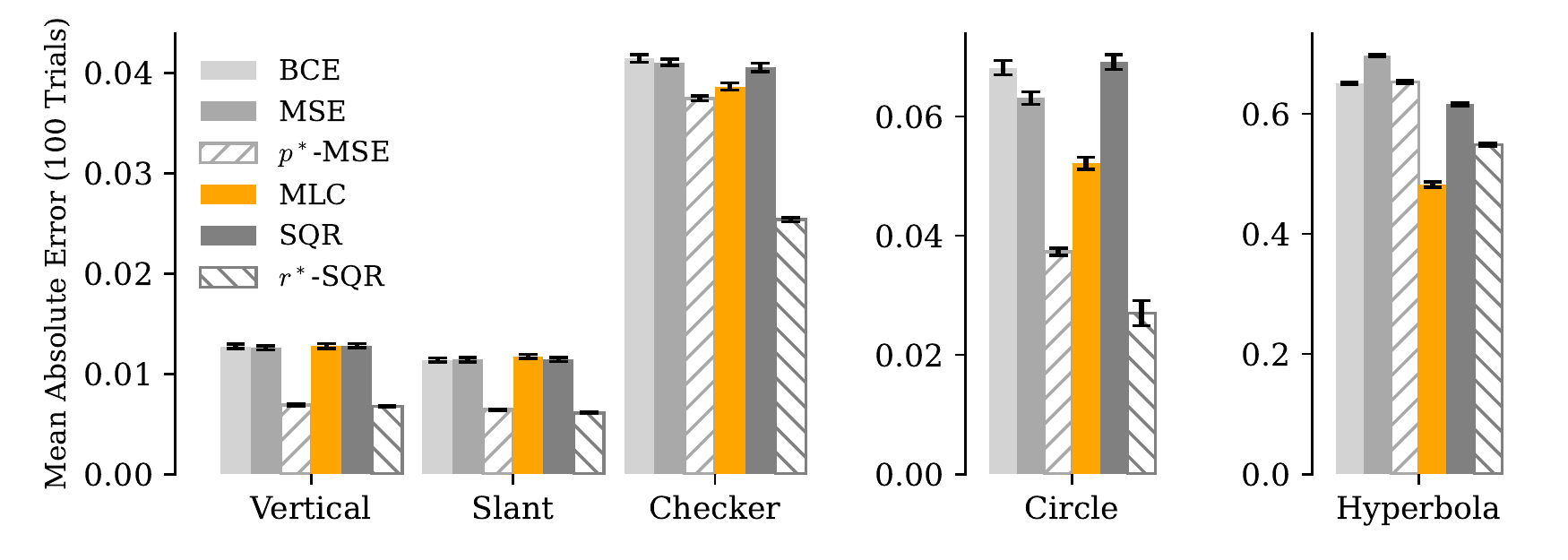}
    \caption{Mean absolute errors are compared for the four different losses considered (binary cross-entropy, mean squared error, maximum likelihood classifier, and square root), each with their respective optimal parametrizations. For each loss, five different multivariate normal cases are studied: ``Vertical'', ``Slant'', ``Circle'', ``Checker'', and ``Hyperbola''. Errors represent the standard deviation across 100 independent model trainings.}
    \label{fig:mvn_best_losses}
\end{figure*}

\subsection{Generalized Loss Families}
\label{section:mvn_families}

\subsubsection{Motivation}
By treating the square in the MSE loss and the root in the SQR loss as parameters $p$ and $r$, respectively, we were able to generalize those loss functional to entire continuous parametric families of losses. We saw in the univariate case that we can optimize over $p$ and $r$, and were even able to see, through examining the landscapes of the different loss functionals in a simple case, what kinds of values of $p$ and $r$ will correspond to ``better" loss functionals.

We now continue this investigation in the situation of multivariate Gaussians to get a sense of how much the trend we observe continues into more complex situations.

\subsubsection{Methods}
We used the same methods as in Section \ref{section:families}; in this case, however, we worked with the five different multivariate Gaussians cases rather than the single univariate Gaussians case.

\subsubsection{Results}
In Table \ref{tab:mvn_pr}, we list the optimal values $p^*$ and $r^*$ in each of the five cases. An overall comparison of the four loss functionals with optimized parameterizations alongside $p^*$-MSE and $r^*$-SQR losses is shown in Figure \ref{fig:mvn_best_losses}. The plots of the MAE for the various values of $p$ and $r$ are presented in Appendix \ref{sec:mvn_appendix}.

\begin{table}[ht]
\def\arraystretch{2} 
\centering
    \begin{tabular}{m{15mm}cc}
    Case            & $p^*$     & $r^*$ \\ \hhline{===}
    Vertical        & 1.12      & 0.018 \\
    Slant           & 1.16      & 0.018 \\
    Circle          & 1.28      & $-0.1$ \\
    Hyperbola       & $-0.44$   & $-0.2$ \\
    Checker         & 1.6       & $-0.1$ 
    \end{tabular}
    \captionof{table}{The optimal values for $p^*$ and $r^*$ for the five different multivariate Gaussian cases. Note that in the univariate Gaussian case, the optimal values chosen were $p^* = 1.24$ and $r^* = 0.018$. \label{tab:mvn_pr}}
\end{table}

\subsubsection{Discussion}
The simpler multivariate cases considered (Vertical, Slant) result in very similar values to those found in the univariate Gaussian case: $p^*$ and $r^*$ are close to 1.24 and 0.018, respectively. 

In the more complex multivariate cases (Circle, Hyperbola, and Checker), the optimal values of $p^*$ are also between 1 and 2, with the expection of Hyperbola, for which $p^* = -0.44$. It's possible that an equally-performing value of $p$ larger than $2$ could also exist, but our studies did not scan far enough to probe the asymptotic behavior in that direction. The optimal values $r^*$ are all negative. However, it is worth noting that the MAE landscapes for the $r$-SQR are symmetric, and the corresponding MAEs for $|r^*|$ are small, so the signs of these values is likely due to random chance. For these cases, the optimal values of $r^*$ are less than 1, as in the univariate case, but very small values of $r^*$ ($|r^*| < 0.01$) are too numerically unstable to consistently yield useful outputs.

Overall, as shown in Figure \ref{fig:mvn_best_losses}, if one chooses only from the four loss functionals as defined in Table \ref{tab:losses}, but with optimized parametrizations, all four show equally good performance for the simpler cases (Vertical, Slant), but the MLC loss is significantly better than the other three choices in the more complex cases (Circle, Hyperbola, Checker). However, if one chooses $p^*$ and $r^*$ by scanning along these generalized loss families, the improvements are immense: for all cases except Hyperbola, the optimized $r^*$-SQR MAE is between 30\% and 50\% smaller than the optimized MLC MAE.
\section{Physics Data}
\label{sec:hep}
\begin{figure*}[ht]
    \includegraphics[width=0.83\linewidth]{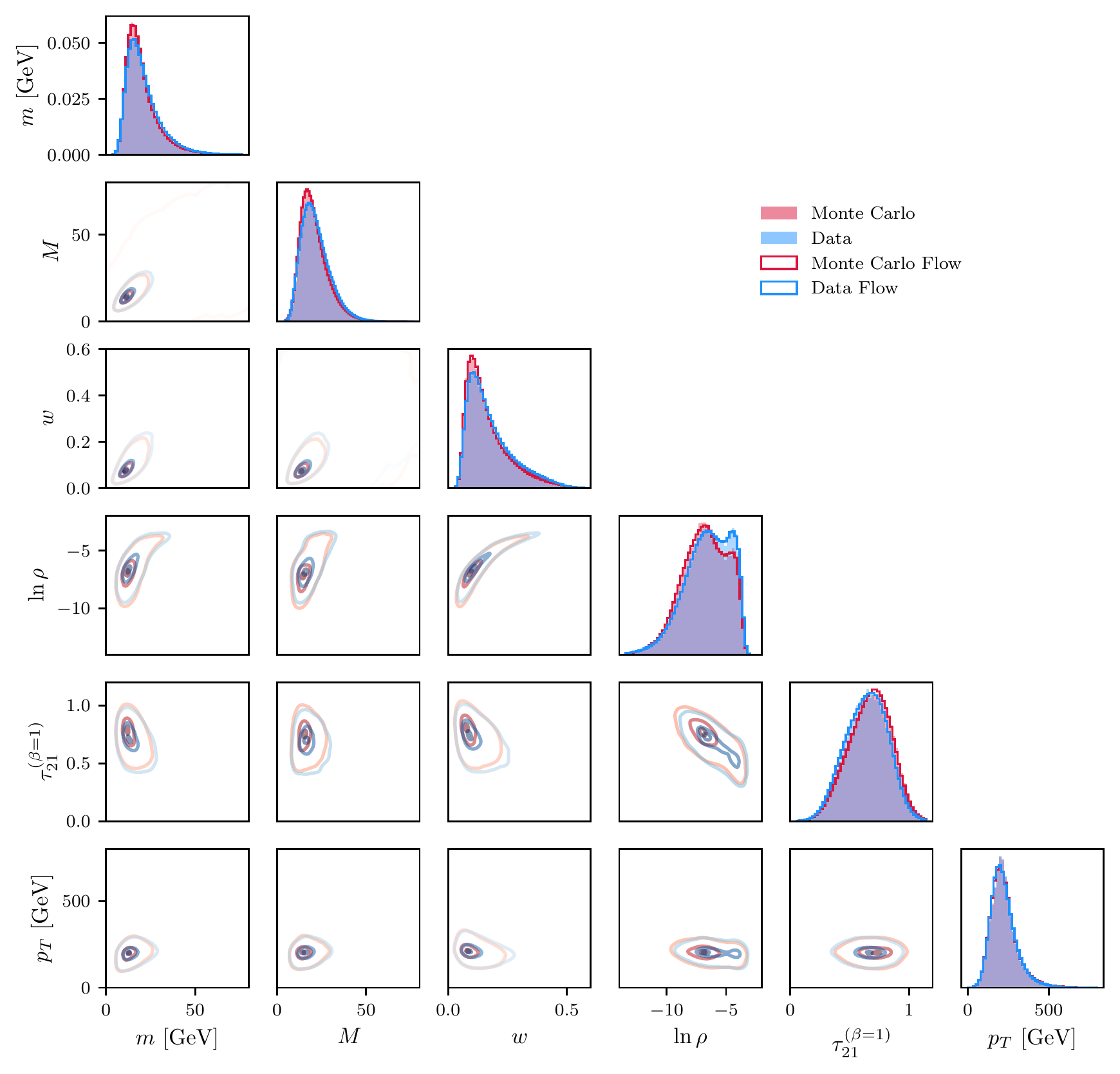}
    \caption{A corner plot indicates the correlations of the six features. Blue and red contours correspond to the particle-level and the detector-level data, respectively.}
    \label{fig:pairs}
\end{figure*}
\subsection{\texorpdfstring{Parametrizing \(f\)}{Parametrizing f}}
\subsubsection{Motivation}
In our final case study, we extended our comparison of classifier parametrizations and loss functionals to simulated high-energy particle physics datasets \cite{zenodo}. While there are a number of observables present in the datasets, in our analysis we considered six observables in total: the jet mass $m$, constituent multiplicity $M$, jet width $w$, the jet mass after Soft Drop grooming $\ln{\rho}= \ln{(m_{SD}^2/p_T^2)}$, the $n$-subjettiness ratio $\tau_{21} = \tau_2^{(\beta = 1)} / \tau_1^{(\beta = 1)}$, and the transverse momentum $p_T$.

The datasets consist of particle-level and detector-level simulated QCD jets originating from $Z$ + jets events. $Z$ + jets events from proton-proton collisions generated at $\sqrt{s} = 14$ TeV were simulated using \textsc{Herwig} 7.1.5 \cite{herwig1, herwig2, herwig3} with the default tune and \textsc{Pythia} 8.243 \cite{pythia1, pythia2, pythia3} tune 21 \cite{pythiatune} (ATLAS A14 central tune with NNPDF2.3LO). We call the \textsc{Pythia} simulation ``Monte Carlo'' and \textsc{Herwig} ``data''. For the generated events, the $p_T$ of the $Z$ boson is required to be larger than 150 GeV. Events then are passed through the \textsc{Delphes} 3.4.2 fast detector simulation \cite{delphes} of the CMS detector. The datasets consist of the highest-momentum jet from $Z$ boson events with $p_T \geq 200$ GeV. This process ultimately yields about 1.6 million jets for each simulation. Figure \ref{fig:pairs} displays histograms of each of the six observables for both the ``Monte Carlo'' and the ``data''.

In this more complex setting, we no longer have access to the true likelihood ratio, as we do not know the underlying distributions generating these datasets. To allow for a more complete comparison of the different parametrizations' ability to model the ``true" likelihood ratio, we therefore fit Normalizing Flows \cite{rezende2016variational} to each sample. These flows estimate the generating distribution of the samples, and thus allow us to compute ``true" likelihood ratios for these datasets.

\subsubsection{Methods}
We first trained a Normalizing Flow \cite{rezende2016variational} for each of the ``Monte Carlo'' and ``data'' simulated samples. All flows were implemented in \textsc{PyTorch} \cite{pytorch} using the \textsc{nflows} package \cite{nflows}. The flow networks consisted of five layers with a layer size of 8, and were trained for \(10^7\) epochs. Each layer consisted of a reverse permutation transform followed by a masked affine autoregressive transform. The flows were optimized to maximize the log likelihood of the data using the \textsc{Adam} \cite{adam} optimizer. For each of the 100 models with the lowest overall loss, an additional classifier was trained to distinguish between events from the flow and the simulated data. The model with the lowest classifier AUC was then selected as a proxy for the underlying distributions of these datasets. New training and testing datasets were then constructed by sampling \(10^7\) and \(10^5\) events, respectively, from the selected flows that fell within the ranges of the relevant simulated observables.

The methodology following this point closely follows the methodology established in Sec.~\ref{section:univariate_methods}. We implemented all four classifiers with each of the three parametrizations on the dataset, training 100 independent copies of each classifier architecture to minimize the corresponding loss functional. We used the MAE as the performance metric, computed as in Equation \ref{eq:mae}. In particular, it was computed with the true likelihood ratio \(\lr(X)\) from the flows and the model likelihood ratio \(\hat{\lr}(X)\) averaged over the 100 copies of each classifier. As before, the MAE was computed as an empirical average over 100,000 samples.

\subsubsection{Results}
The distributions of the ``data'' and ``Monte Carlo'' learned by the flows are plotted along side the empirical distributions in Figure \ref{fig:pairs}. To quantify the quality of the flows' learned distributions, we trained classifiers to try to distinguish between proxy datasets sampled from the flows and the original datasets; for the ``Monte Carlo'', the AUC was 0.5094, and for the ``data'', the AUC was 0.5100. These AUCs close to 0.5 indicate that the classifier has difficulty distinguishing between these two distributions, and therefore that the flows have performed reasonably well at reflecting the target distributions.

\begin{figure*}[h]
    \centering
    \begin{subfigure}[b]{0.6\textwidth}
        \centering
    \includegraphics{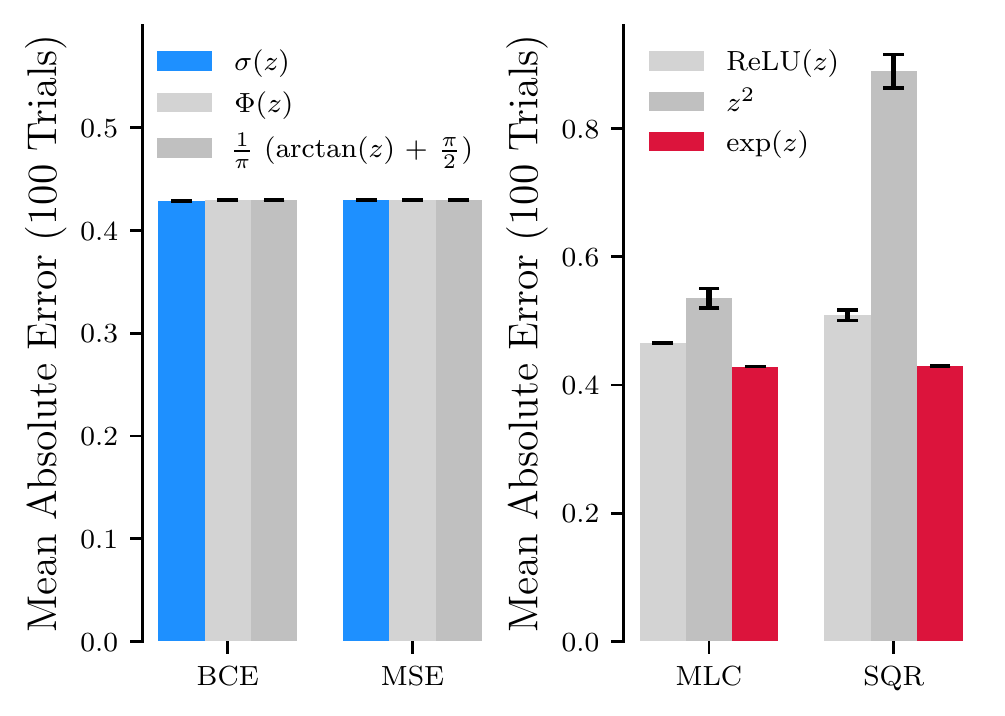}
    \caption{Comparing Parametrizations}
    \end{subfigure}
    \begin{subfigure}[b]{0.35\textwidth}
        \centering
    \includegraphics[width=\textwidth]{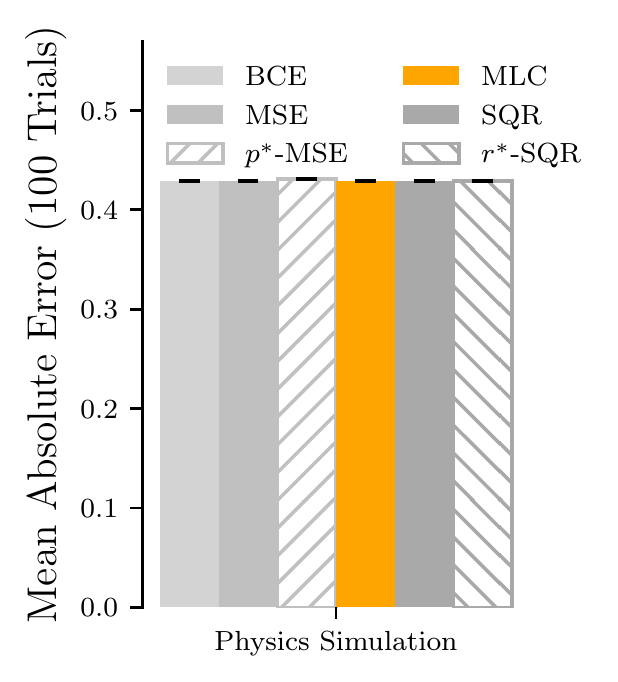}
        \caption{Comparing Optimized Losses}
    \end{subfigure}
    \caption{The MAEs are compared for the \textsc{Pythia}/\textsc{Herwig} + \textsc{Delphes} particle physics jet datasets \cite{zenodo} for the four different losses considered. Errors represent the standard deviation across 100 independent model trainings. In \textbf{(a)}, each loss is shown with 3 different parametrizations. In \textbf{(b)}, the best-performing parametrization is chosen for each loss, and these optimized losses are then directly compared.}
    \label{fig:physics_losses}
\end{figure*}

\begin{figure*}
    \includegraphics{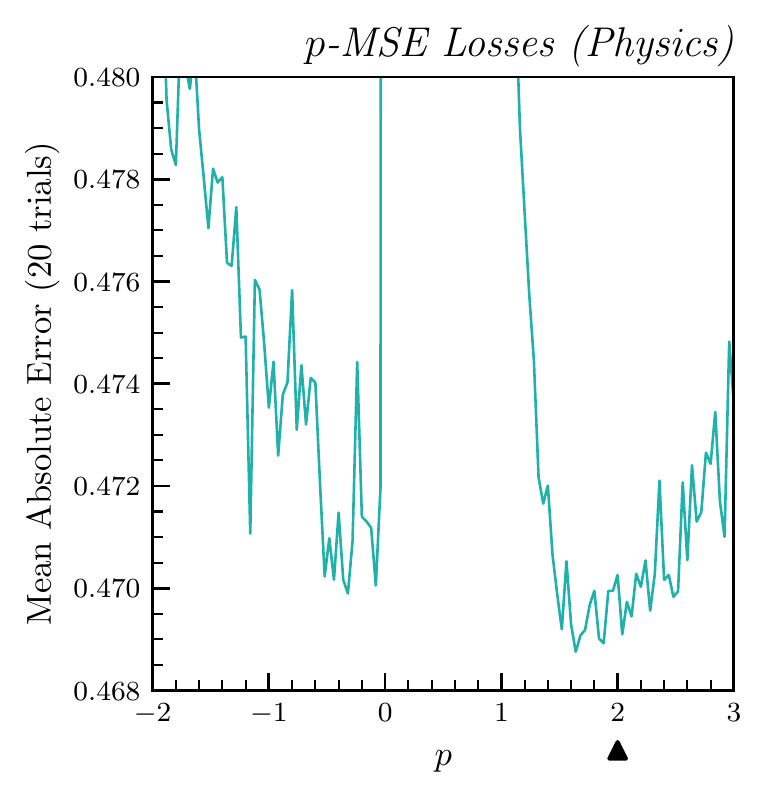}
    \includegraphics{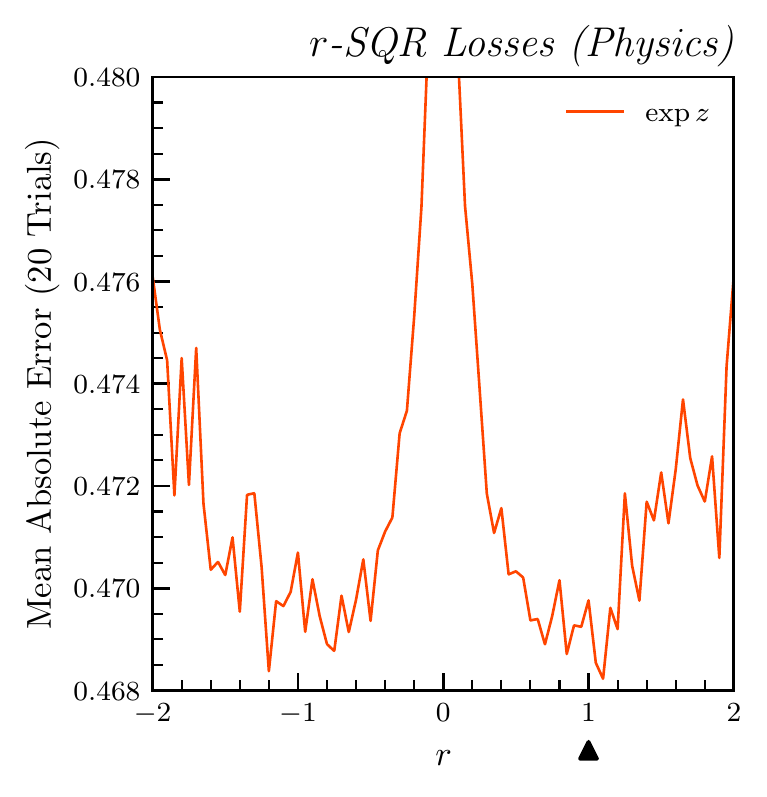}
    \caption{(a) The mean absolute errors averaged over logistically-parametrized models trained with the generalized MSE loss family. The mean absolute error is minimized at \(p^* = 1.64\). The arrow indicates the typical choice of $p=2$ for the standard MSE loss. (b) The mean absolute errors averaged over exponentially-parametrized models trained with the generalized SQR loss family. The mean absolute error was smallest at \(r^* = 1.1\). The arrow indicates the typical choice of $r=1$ for the standard SQR loss.}
    \label{fig:physics_rp_landscapes}
\end{figure*}

\subsubsection{Discussion}
The observed trend in the Gaussian studies that the logistical and exponential parametrizations were the best for the BCE/MSE and MLC/SQR losses, respectively, also holds in the physics case, as shown in Figure \ref{fig:physics_losses}. Of the four optimized loss functionals, the MLC loss with exponential parametrization performs better than the other three loss configurations.

\subsection{Generalized Loss Families}
\subsubsection{Motivation}
In the previous studies with univariate and multivariate Gaussians, we found that the performances of likelihood ratio models trained with losses from the generalized families of $p$-MSE and $r$-SQR losses followed a similar structure across various cases. In order to examine the robustness of this observed structure, we repeated the same study with the high energy particle physics dataset.

\subsubsection{Methods}
The methodology for this study was similar to that of the previous studies. We scanned over values of $p$ and $r$ in the intervals $[-2, 3]$ and $[-2, 2]$. For each increment of $p$, we trained 20 models with the $p$-MSE loss functional defined by that value of $p$ and averaged together their mean absolute errors. Likewise, for each increment of $r$, we trained 20 models with the $r$-MSE loss functional defined by that value of $r$ and averaged together their mean absolute errors. We parametrized the $p$-MSE classifiers with logistic activation functions and the $r$-SQR classifiers with exponential activation functions. All models were trained on the same set of one million samples from the flows fit to the distributions of the physics data.

\subsubsection{Results}
The plots of the MAEs of the likelihood ratio models for the loss functionals are provided in Figure \ref{fig:physics_rp_landscapes}. As before, we observe vertical features in the plots when the loss functional is no longer strictly convex ($p \in (0, 1)$ and $r = 0$). The MAE was minimized at $p^* = 1.64$ and $r^* = 1.1$. A comparison of the $p^*$-MSE and $r^*$-SQR losses with these chosen values alongsize the other four losses with optimized parametrizations is shown in Figure \ref{fig:physics_losses} and summarized in Table~\ref{tab:physics}.

\begin{table}[ht]
\centering
\begin{tabular}{m{12mm}|cc|cc|cc}
Loss       & $\sigma(z)$   & Error     & $\Phi(z)$    & Error     & $\text{tan}^{-1}(z)$    & Error     \\ \hhline{=|==|==|==}
BCE             & \textbf{0.4291}        & 0.0002    & 0.4300    & 0.0003    & 0.4294    & 0.0002    \\
MSE             & \textbf{0.4294}        & 0.0002    & 0.4297    & 0.0003    & 0.4297    & 0.0002    \\
$p^*$-MSE       & \textbf{0.4309}        & 0.0002    &---        & ---       & ---       & ---       \\
&&&&&& \\
Loss   & ReLU$(z)$     & Error     & $z^2$     & Error     & $e^z$     & Error\\ \hhline{=|==|==|==}
MLC         & 0.4656        & 0.0006    & 0.5351    & 0.0153    & \textbf{0.4287}    & 0.0002 \\
SQR         & 0.5086        & 0.0082    & 0.8896    & 0.0262    & \textbf{0.4294}    & 0.0002   \\
$r^*$-SQR   & ---           & ---       &---        & ---       & \textbf{0.4291}    & 0.0002  \\
\end{tabular}
\vspace*{0.2in}
\captionof{table}{Mean absolute errors are computed for various loss functional configurations in the classification of two simulated high-energy physics datasets. 100 independent and identical classifiers were trained for each configuration to calculate the uncertainties. Errors represent one standard deviation. The activation functions in the first column represent the typical choices for each loss functional. \label{tab:physics}}
\end{table}

\subsubsection{Discussion}
The shape of the $p$ and $r$ scans looks approximately similar to those observed in the previous case studies (e.g. Figure \ref{fig:mse_sqr_ab}); however, since the MAE landscape is flat away from the non-convex regions ($p \in (0, 1)$ for $p$-MSE and $r = 0$ for $r$-SQR), the best choices $p^* = 1.64$ and $r^* = 1.1$ perform about the same as the unoptimized choices of $p = 2$ and $r = 1$. In this particular case, the evidence does not suggest that changing $p^*$ or $r^*$ from their default values of 2 and 1, respectively, would yield a significant benefit in reducing the mean absolute error. It is possible that better values exist beyond the ranges of $r, p$ considered here. Overall, as shown in Figure \ref{fig:physics_losses} and  Table~\ref{tab:physics}, the best-performing loss is MLC with exponential parameterization. Pairing the MLC loss with a suboptimal activation function, however, introduces additional penalties for the MAE compared with the optimal choice of $e^z$: $+8\%$ for $\text{ReLU}(z)$ and $+20\%$ for $z^2$. 

\section{Conclusions}
\label{sec:conclusions}
The likelihood ratio $\lr(x) = \frac{p(x \mid \theta_0)}{p(x \mid \theta_1)}$ is a statistical quantity essential for characterizing whether an experimental dataset $x$ better supports one of two hypotheses defined by sets of parameters $\theta_0$ and $\theta_1$. It is used beyond hypothesis testing, too, for applications such as reweighting high-dimensional distributions for background estimation and more. In contexts where calculating the likelihood ratio is impossible or very tedious, researchers can use the ``likehood ratio trick'', leveraging a neural network classifier to approximate the likelihood ratio. 

Often, the likelihood ratio trick is implemented by minimizing a typical choice of loss functional for a classifier: the binary cross-entropy loss. However, many loss functionals satisfy the likelihood ratio trick setup. 

In this paper, we presented detailed studies comparing four choices of loss functionals: 
binary cross-entropy (BCE), mean squared error (MSE), maximum likelihood classifier (MLC), and square root (SQR). For each of these four loss functionals, we also explored a suite of choices of final activation functions for parametrizing the neural network output. For the MSE and SQR losses, we performed a scan along the exponential parameter (replacing $2\rightarrow p$ for MSE and replacing $\frac{1}{2}\rightarrow \frac{r}{2}$ for SQR) to understand the behavior of these generalized families of loss functionals. 

\enlargethispage{\baselineskip}

As a result of these studies, we present the following recommendations for optimized implementations of each of these loss functionals in the likelihood ratio trick:

\begin{center}
\def\arraystretch{1.5} 
\begin{tabular}{m{65mm}cc}
   Loss  & Activation \\ \hhline{===}
   Binary Cross-Entropy (BCE)  & $\sigma(z)$ \\
   Mean Squared Error (MSE) & $\sigma(z)$ \\
   Maximum Likelihood Classifier (MLC) & exp$(z)$ \\ 
   Square Root (SQR) & exp$(z)$
\end{tabular}
\end{center}

For MLC and SQR losses, we find that choosing small, nonzero values of $r$ (and, correspondingly, $p = r + 1$) tend to result in smaller mean absolute errors than the default choices ($r=1$ and $p=2$) for these loss functionals. As we illustrate by mapping the loss landscape of a simple neural network, this is because smaller values of $r$ can yield shallower loss landscapes where many values are nearly optimal, while larger values of $r$ have steeper landscapes for models to traverse, with a much smaller proportion of the phase space corresponding to optimum values of the loss. 

The loss landscape will vary with each new application, so we recommend that future researchers perform a scan along $p$ or $r$ to find an optimum value as part of hyperparameter optimization. If a scan over $p$ or $r$ is not feasible, we recommend comparing the default selections (i.e. $p=2$ and $r=1$) with our alternative recommendations derived from the average optimum values across our various trials $p^*=1.25$ and $r^*=0.1$, or:

\begin{align*}
L_{\text{MSE}^*}[f] &= -\int \text{d}x \bigg((1-f)^{1.25}p(x \mid \theta_0)\\
&\phantom{= -\int\d{x} \bigg(}+ f^{1.25}p(x \mid \theta_1)\bigg)\\
L_{\text{SQR}^*}[f] &= -\int \text{d}x \bigg(f^{-0.05}p(x \mid \theta_0) + f^{0.05}p(x \mid \theta_1)\bigg).
\end{align*}

Across the majority of the various datasets we considered, these choices tend to have significantly smaller mean absolute errors than the default selections while maintaining good numerical stability across multiple trainings. An interesting future investigation would be to consider how to dynamically optimize $p$ and $r$ as learned parameters during training. 

When tested on univariate Gaussians and simple multivariate Gaussians (Vertical and Slant cases), all four loss implementations with optimized parametrizations perform similarly when approximating the desired likelihood ratio. For larger datasets ($N > 10^5$), choosing different exponents in the definitions of MSE and SQR loss functionals results in an additional $\geq50$\% reduction in errors for these cases.

On more complex datasets, including multidimensional Gaussians (Checker, Hyperbola, Circle) as well as simulated high-energy physics data, the Maximum Likelihood Classifier (MLC) loss with exponential parametrization performs the best out of the four default losses considered. Compared with this choice, other combinations of loss functionals and activation functions saw increased MAE values of between 0.5\% and 82\%. Choosing different exponents in the definitions of MSE and SQR loss functionals additionally results in between 30\% and 50\% smaller errors for the Checker and Circle cases. For the Hyperbola and simulated high-energy physics case, choosing alternate $p^*$ and $r^*$ values in the range $[-2,3]$ does not yield a significant performance improvement, though it is possible that better values could exist outside of this range. Overall, the results are clear that the choice of loss functional and activation function is one that should be taken carefully, as there are potentially large penalties for the MAE associated with a suboptimal selection.

While these configurations performed well in our chosen case studies, these results should not be read as a guarantee that these choices will result in optimal performance for any dataset. We therefore recommend that other researchers compare the results of several of the optimized losses described in this work to yield the most effective setup for a given dataset. 

There remain several open questions in this line of inquiry. For instance, can an analytical analysis of these loss functionals explain some of the performance differences observed? How much can we further characterize the uncountably many possible loss functionals that satisfy this setup? How else can we generalize certain loss functionals? Further investigations could include a more involved parametrization of the loss functionals, such as by splines, which covers more loss functionals than the ones explored by the $p$ and $r$ scans here. Pursuing these threads can help achieve even better scientific measurements enabled by machine learning in the near future.

\section*{Code Availability}
A codebase with instructions on how to reproduce each of the plots in this paper is located at \url{https://github.com/shahzarrizvi/reweighting-schemes}.

\section*{Acknowledgements}
 We are grateful to Jesse Thaler for very helpful feedback about figures-of-merit and ways to generalize our loss functions. We thank Dag Gillberg for the suggestion to compare NNs with BDTs and Vinicius Mikuni for the idea of using normalizing flows to model the LR of the physics datasets. We thank Lindsey Gray for his idea of exploring the space of loss functionals via parametrizations with splines. M.P. thanks Shirley Ho and the Flatiron Institute for their hospitality while preparing this paper. S.R., M.P., and B.N. are supported by the Department of Energy, Office of Science under contract number DE-AC02-05CH11231.

\bibliography{bib}{}
\clearpage

\appendix

\section{Non-Gaussian Distributions}
\label{sec:nongaussian}
As an extension to our studies on univariate Gaussians in Section \ref{section:univariate_methods}, we also examined two non-Gaussian univariate distributions: Beta and Gamma distributions. 

\begin{figure*}[h]
    \centering
    \begin{subfigure}[b]{0.48\linewidth}
        \centering
        \includegraphics{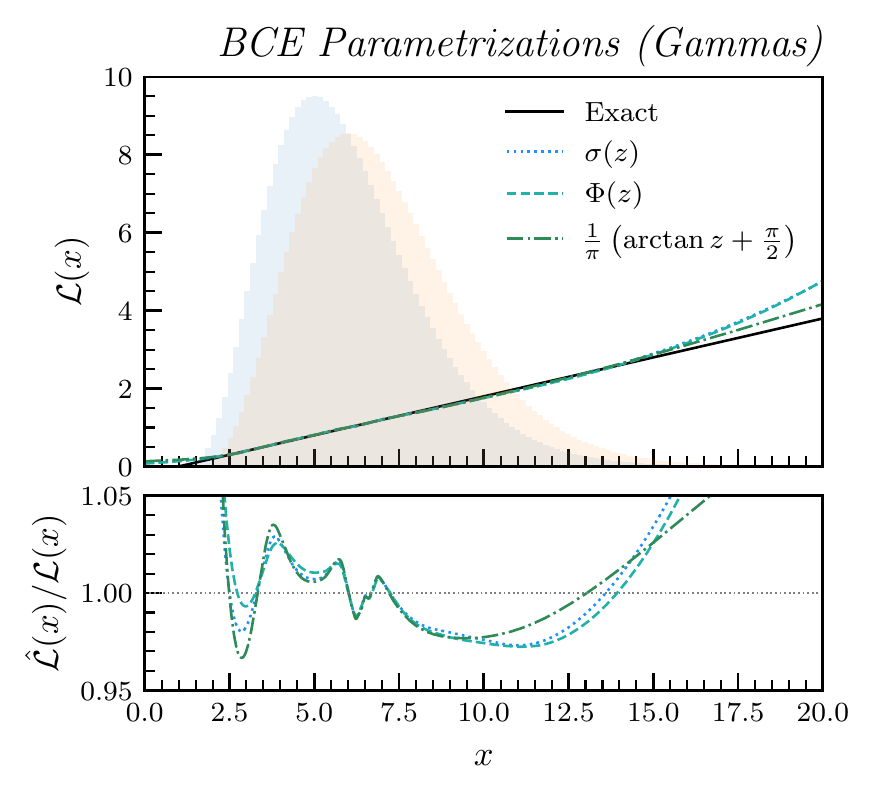}
        \caption{}
    \end{subfigure}
    \begin{subfigure}[b]{0.48\linewidth}
        \centering
        \includegraphics{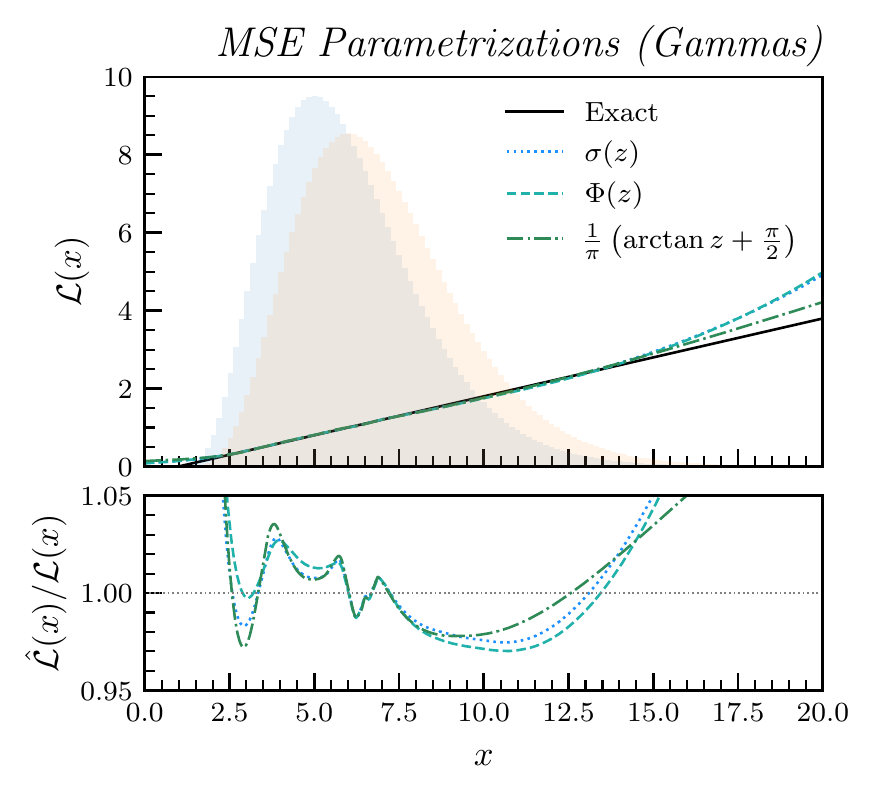}
        \caption{}
    \end{subfigure}
    \caption{Parametrizations of $f$ for the BCE and MSE losses. (a) The average likelihood ratio fits of the logistic, Gaussian CDF, and arctangent parametrizations for the BCE loss, with mean absolute errors 0.0168, 0.0179, and 0.0162, respectively. (b) The average likelihood ratio fits of the logistic, Gaussian CDF, and arctangent parametrizations for the MSE loss, with mean absolute errors 0.0170, 0.0196, and 0.0161, respectively.}
    \label{fig:gamma_bce_mse}
\end{figure*}

\begin{figure*}[h]
    \centering
    \begin{subfigure}[b]{0.48\linewidth}
        \centering
        \includegraphics{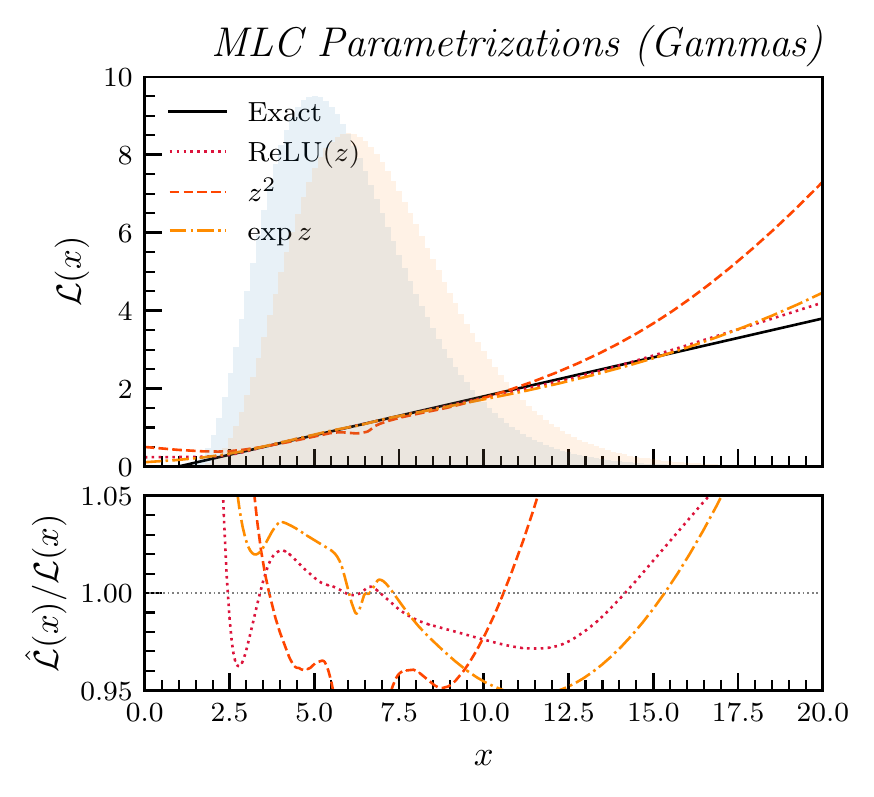}
        \caption{}
    \end{subfigure}
    \begin{subfigure}[b]{0.48\linewidth}
        \centering
        \includegraphics{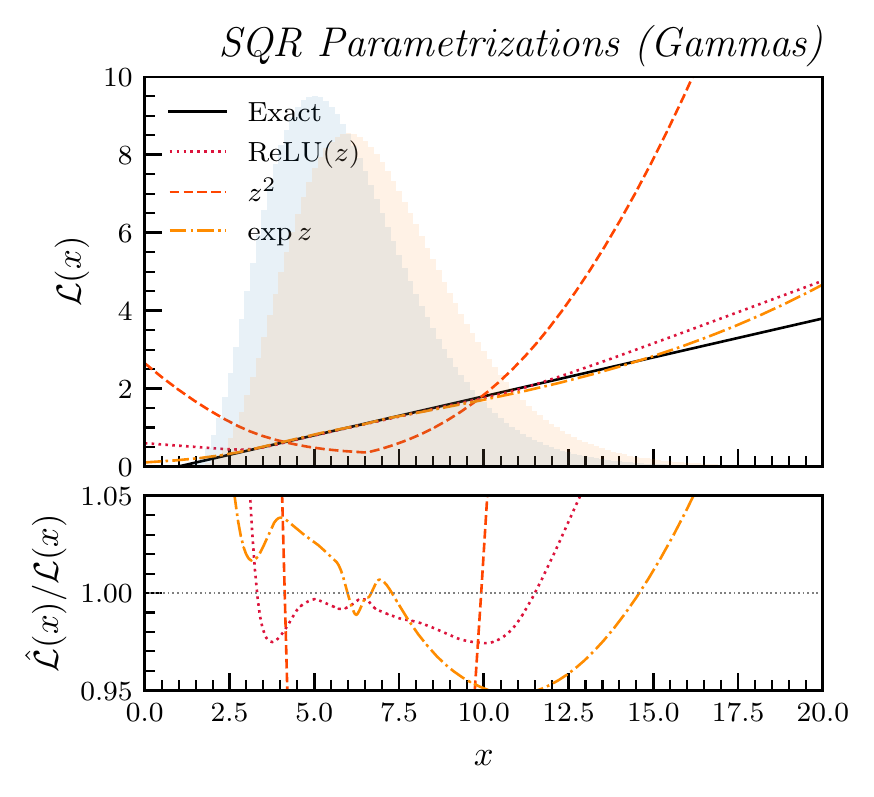}
        \caption{}
    \end{subfigure}
    \caption{Parametrizations of $f$ for the MLC and SQR losses. (a) The average likelihood ratio fits of the ReLU, square, and exponential parametrizations for the MLC loss, with mean absolute errors 0.0165, 0.126, and 0.0275, respectively. (b) The average likelihood ratio fits of the ReLU, square, and exponential parametrizations for the SQR loss, with mean absolute errors 0.0343, 0.761, and 0.0285, respectively.}
    \label{fig:gamma_mlc_sqr}
\end{figure*}

\begin{figure*}[h]
    \centering
    \begin{subfigure}[b]{0.6\textwidth}
        \centering
    \includegraphics{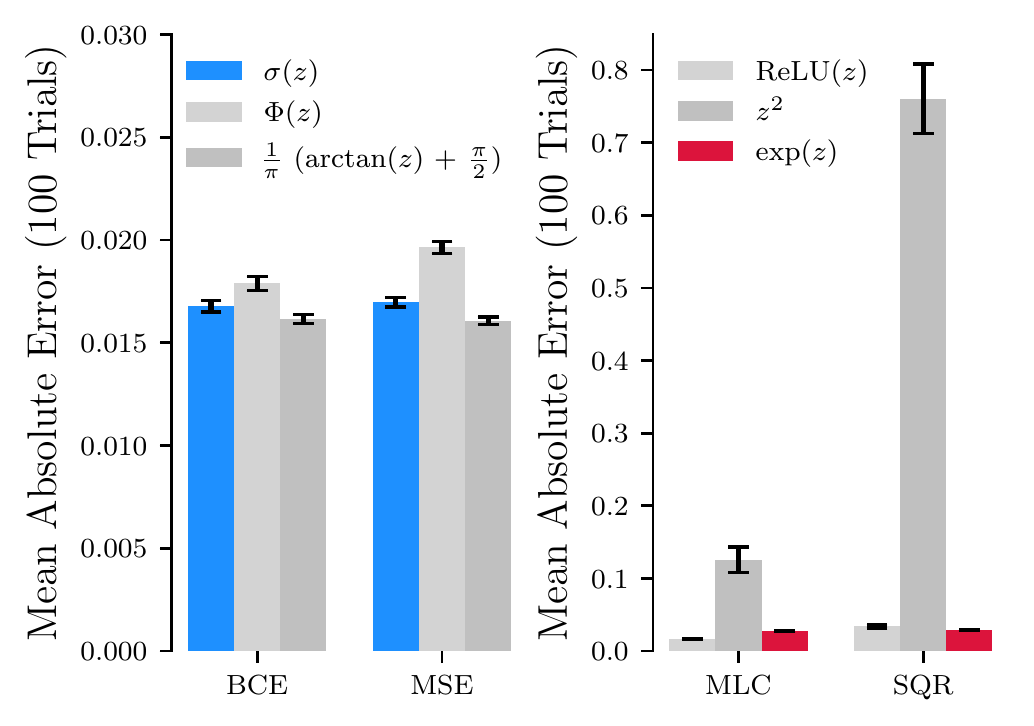}
    \caption{Comparing Parametrizations}
    \end{subfigure}
    \begin{subfigure}[b]{0.35\textwidth}
        \centering
    \includegraphics[width=\textwidth]{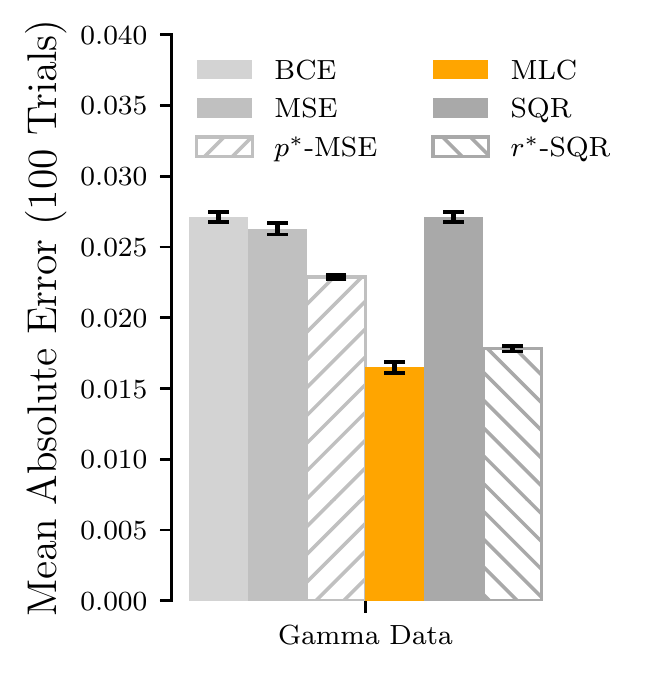}
        \caption{Comparing Optimized Losses}
    \end{subfigure}
    \caption{The MAEs are compared for gamma-distributed data for the four different losses considered. Errors represent the standard deviation across 100 independent model trainings. In \textbf{(a)}, each loss is shown with 3 different parametrizations. In \textbf{(b)}, the best-performing parametrization is chosen for each loss, and these optimized losses are then directly compared.}
    \label{fig:gamma_losses}
\end{figure*}

Our first case study considered two Gamma distributions with different shape parameters and identical unit rates: \(X_0 \sim {\rm Gamma}(6, 1)\) and \(X_1 \sim {\rm Gamma}(5, 1)\). We conducted the same parametrization study on them as in Section \ref{sec:univar_b}. The resulting likelihood ratio approximations are displayed in Figure \ref{fig:gamma_bce_mse} and \ref{fig:gamma_mlc_sqr}. Figure \ref{fig:gamma_losses}a) directly compares the performances of the different parametrizations of each of the four loss functionals.

For the MSE and BCE loss functionals, the arctangent parametrization performed the best, but only by a small amount; the next best for both was the logistic parametrization, with an MAE less than \(6\%\) greater than the arctangent parametrization's MAE. The performance for the SQR loss aligned with previous observations: the square parametrization performed worst, followed by the ReLU parametrization; and the exponential performed the best. Interestingly, for the MLC loss, the ReLU parametrization performed noticeably better than the exponential parametrization, with about a 40\% reduction in MAE. This is likely due to the fact that for Gamma-distributed data, the true likelihood ratio is a linear function: $\lr(x) = \frac{1}{5}x$; as a result, the better performance of the ReLU parametrization here for the MLC loss is likely due to inductive bias. 

\begin{figure*}[ht]
    \centering
    \begin{subfigure}[b]{0.48\linewidth}
        \centering
        \includegraphics{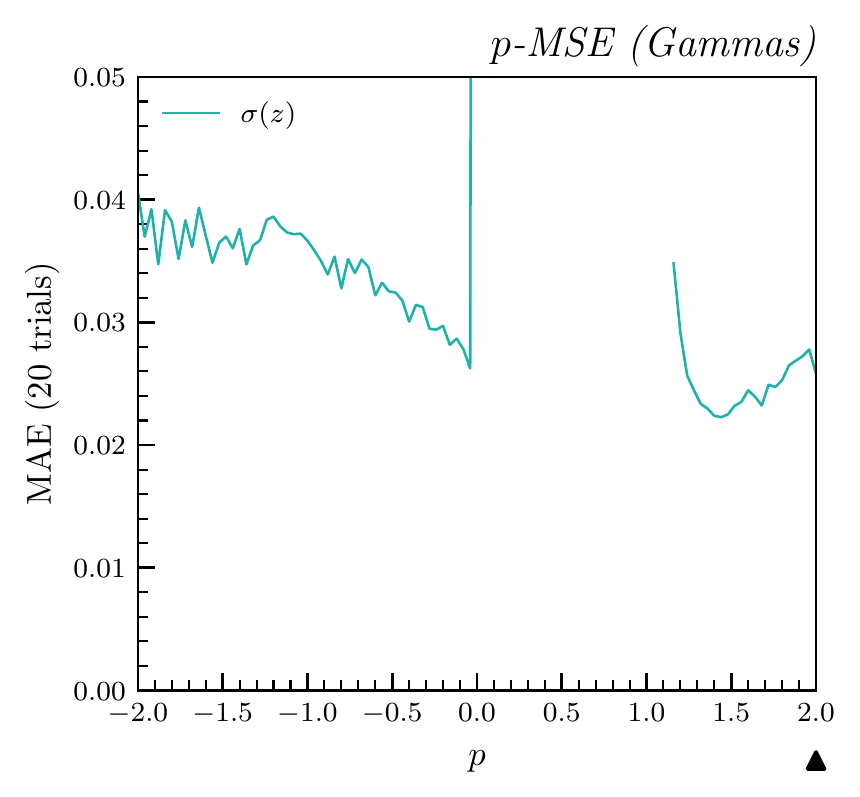}
        \caption{}
    \end{subfigure}
    \begin{subfigure}[b]{0.48\linewidth}
        \centering
        \includegraphics{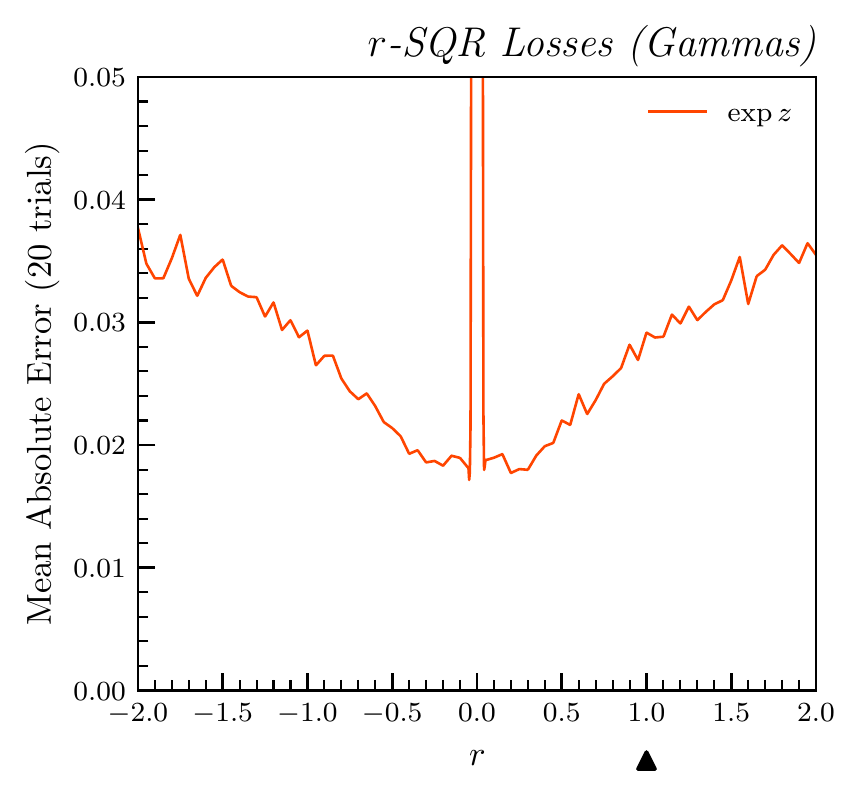}
        \caption{}
    \end{subfigure}
    \caption{(a) The mean absolute errors averaged over models trained on the generalized MSE loss family for the logistic parametrization with gamma-distributed data. The mean absolute error is minimized at \(p^* = 1.44\). The arrow indicates the typical choice of $p=2$ for MSE loss. (b) The mean absolute errors averaged over models trained on the generalized SQR loss family for the exponential parametrization with gamma-distributed data. The mean absolute error was smallest at \(r^* = -0.046\). The arrow indicates the typical choice of $r=1$ for SQR loss.}
    \label{fig:gamma_rp}
\end{figure*}

In addition, we examined the performance of the generalized $p$-MSE And $r$-SQR losses for learning the likelihood ratio of the Gamma distributions. The methodology was identical to that of \ref{section:families}. Figure \ref{fig:gamma_rp} plots the MAE averaged over 20 models for models trained with values of $p$ (for $p$-MSE) and $r$ (for $r$-SQR) in the interval $[-2, 2]$. We found optimal values of $p^* = 1.44$ and $r^* = -0.046$. These results align with our other observations that values of $p$ slightly above $1$ and values of $r$ close to 0 work well for attaining a good likelihood ratio fit.

Figure \ref{fig:gamma_losses}b) compares the best performing parametrization of the losses against each other: the arctangent parametrizations for the BCE and MSE losses, the ReLU parametrization for the MLC loss, and the exponential parametrization for the SQR loss. Also included are the $p^*$-MSE and $r^*$-SQR losses. The overall best-performing loss is the MLC loss with the ReLU parametrization, followed closely by the $r^*$-SQR loss.

\begin{figure*}[h]
    \centering
    \begin{subfigure}[b]{0.48\linewidth}
        \centering
        \includegraphics{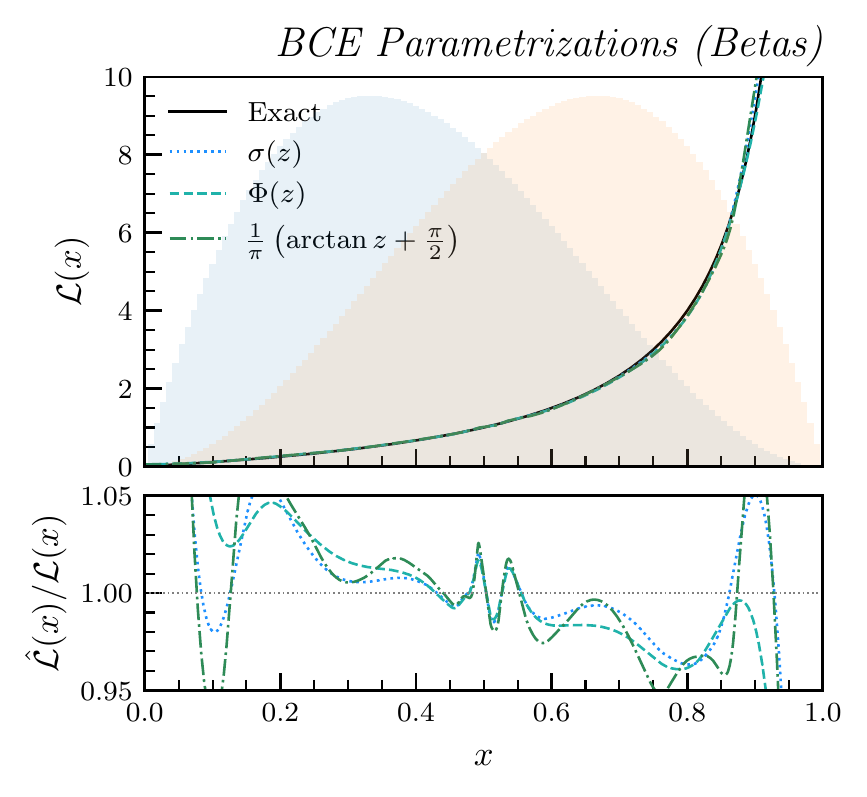}
        \caption{}
    \end{subfigure}
    \begin{subfigure}[b]{0.48\linewidth}
        \centering
        \includegraphics{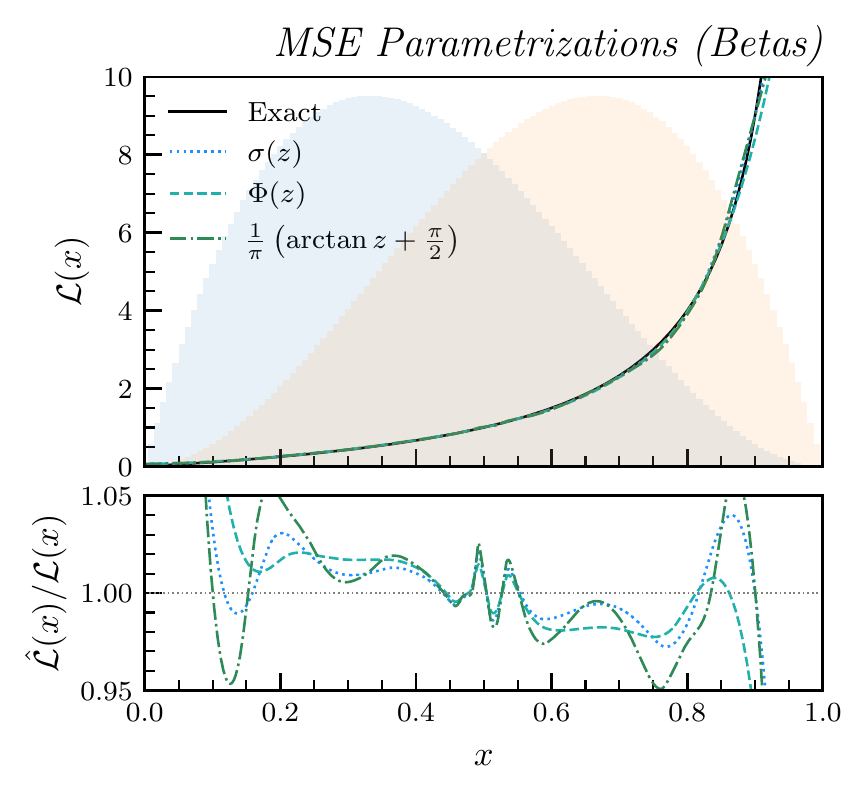}
        \caption{}
    \end{subfigure}
    \caption{Parametrizations of $f$ for the BCE and MSE losses. (a) The average likelihood ratio fits of the logistic, Gaussian CDF, and arctangent parametrizations for the BCE loss, with mean absolute errors 0.220, 0.250, and 0.257, respectively. (b) The average likelihood ratio fits of the logistic, Gaussian CDF, and arctangent parametrizations for the MSE loss, with mean absolute errors 0.259, 0.285, and 0.302, respectively.}
    \label{fig:beta_bce_mse}
\end{figure*}

\begin{figure*}[h]
    \centering
    \begin{subfigure}[b]{0.48\linewidth}
        \centering
        \includegraphics{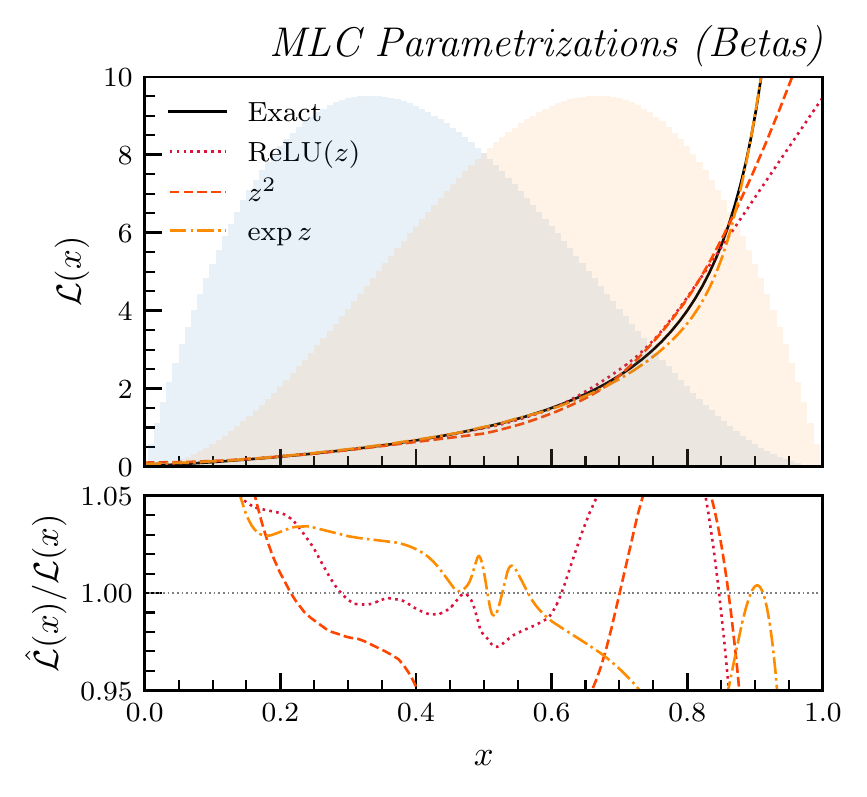}
        \caption{}
    \end{subfigure}
    \begin{subfigure}[b]{0.48\linewidth}
        \centering
        \includegraphics{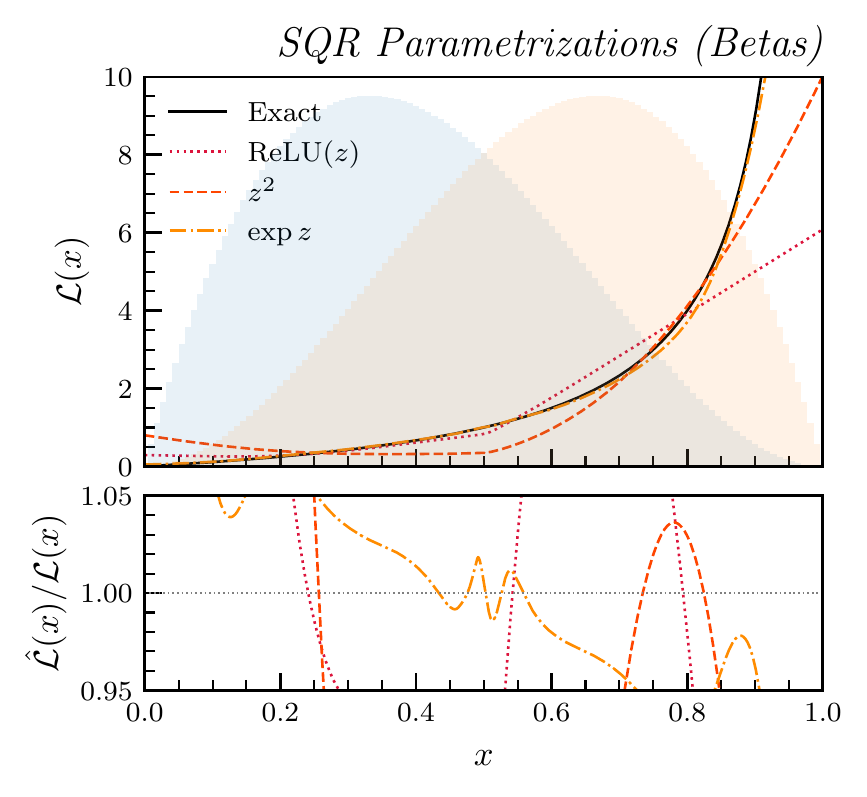}
        \caption{}
    \end{subfigure}
    \caption{Parametrizations of $f$ for the MLC and SQR losses. (a) The average likelihood ratio fits of the ReLU, square, and exponential parametrizations for the MLC loss, with mean absolute errors 0.452, 0.466, and 0.257, respectively. (b) The average likelihood ratio fits of the ReLU, square, and exponential parametrizations for the SQR loss, with mean absolute errors 0.674, 1.171, and 0.288, respectively.}
    \label{fig:beta_mlc_sqr}
\end{figure*}

\begin{figure*}[h]
    \centering
    \begin{subfigure}[b]{0.6\textwidth}
        \centering
    \includegraphics{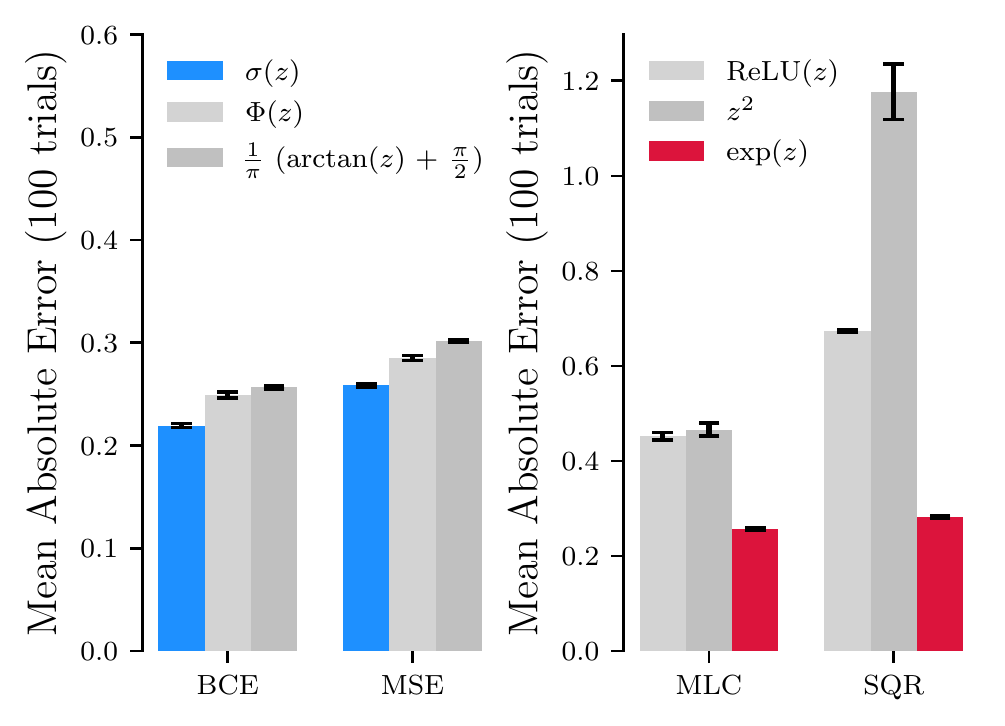}
    \caption{Comparing Parametrizations}
    \end{subfigure}
    \begin{subfigure}[b]{0.35\textwidth}
        \centering
    \includegraphics[width=\textwidth]{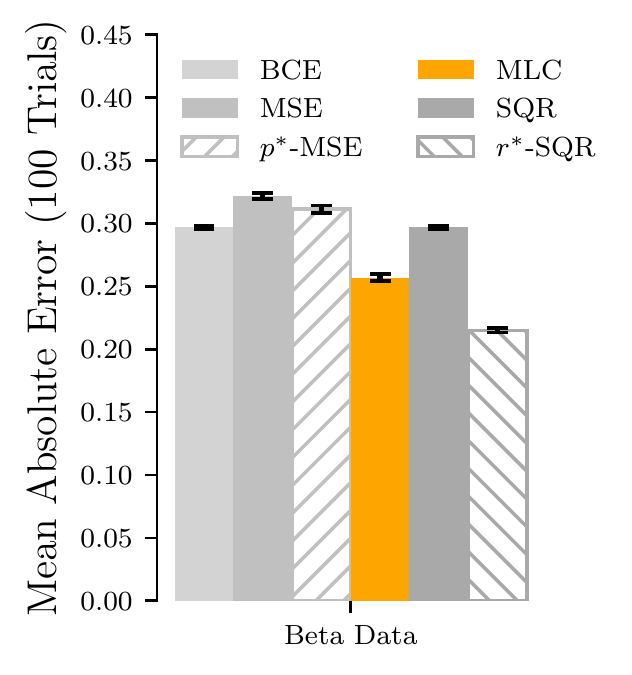}
        \caption{Comparing Optimized Losses}
    \end{subfigure}
    \caption{The MAEs are compared for gamma-distributed data for the four different losses considered. Errors represent the standard deviation across 100 independent model trainings. In \textbf{(a)}, each loss is shown with 3 different parametrizations. In \textbf{(b)}, the best-performing parametrization is chosen for each loss, and these optimized losses are then directly compared.}
    \label{fig:beta_losses}
\end{figure*}

Our final univariate case study considered two different Beta distributions: \(X_0 \sim {\rm Beta} (3, 2)\) and \(X_1 \sim {\rm Beta} (2, 3)\). We conducted the same parametrization study on them as in Section \ref{sec:univar_b}. The resulting parametrization fits are displayed in Figure \ref{fig:beta_bce_mse} and \ref{fig:beta_mlc_sqr}. Figure \ref{fig:beta_losses}a) compares the performances of the different parametrizations of each of the four loss functionals.

In this case study, the best performing classifier parametrization for the BCE and MSE losses was the logistic parametrization, outperforming the other parametrizations by at least \(13\%\). For both the MLC and SQR losses, the exponential parametrization performed the best.

\begin{figure*}[ht]
    \centering
    \begin{subfigure}[b]{0.48\linewidth}
        \centering
        \includegraphics{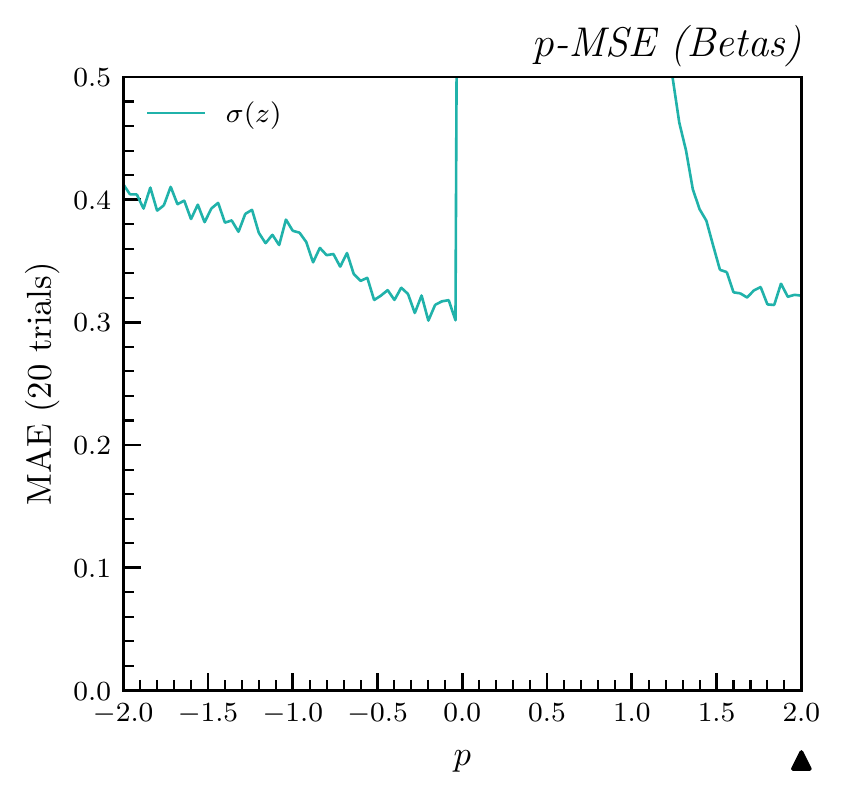}
        \caption{}
    \end{subfigure}
    \begin{subfigure}[b]{0.48\linewidth}
        \centering
        \includegraphics{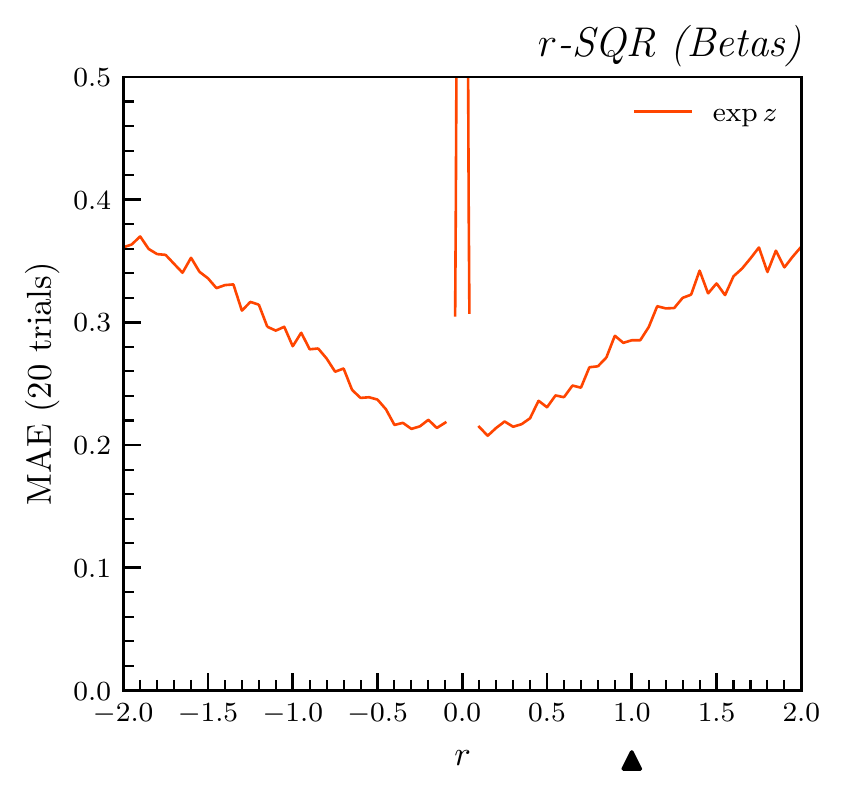}
        \caption{}
    \end{subfigure}
    \caption{(a) The mean absolute errors averaged over models trained on the generalized MSE loss family for the logistic parametrization with beta-distributed data. The mean absolute error is minimized at \(p^* = -0.2\). The arrow indicates the typical choice of $p=2$ for MSE loss. (b) The mean absolute errors averaged over models trained on the generalized SQR loss family for the exponential parametrization with beta-distributed data. The mean absolute error was smallest at \(r^* = 0.15\). The arrow indicates the typical choice of $r=1$ for SQR loss.}
    \label{fig:beta_rp}
\end{figure*}

We also examined the performance of the generalize $p$-MSE And $r$-SQR losses for learning the likelihood ratio of our Beta distributions. The methodology was identical to that of \ref{section:families}. Figure \ref{fig:beta_rp} plots the MAE averaged over 20 models for models trained with values of $p$ (for $p$-MSE) and $r$ (for $r$-SQR) in the interval $[-2, 2]$. We found that $p^* = -0.2$ and $r^* = 0.15$. The value of $r^*$ is consistent with our observation that values of $r$ close to 0 work well. While the value of $p^*$ does not align with the recommendation of values of $p$ slightly above $1$, it is similar to the values of $p^*$ found for some of the more difficult cases we studied, such as with the Hyperbola case study or with the simulated high-energy physics data.

Figure \ref{fig:beta_losses}b) compares the BCE and MSE losses with a logistically-parametrized model, the MLC and SQR losses with exponentially-parametrized models, and the $p^*$-MSE and $r^*$-SQR losses all against one another. The best performing loss was the $r^*$-SQR loss, followed by the MLC loss with an exponential parametrization.

\section{Multivariate Gaussians}
\label{sec:mvn_appendix}
\begin{figure*}[ht]
    \includegraphics{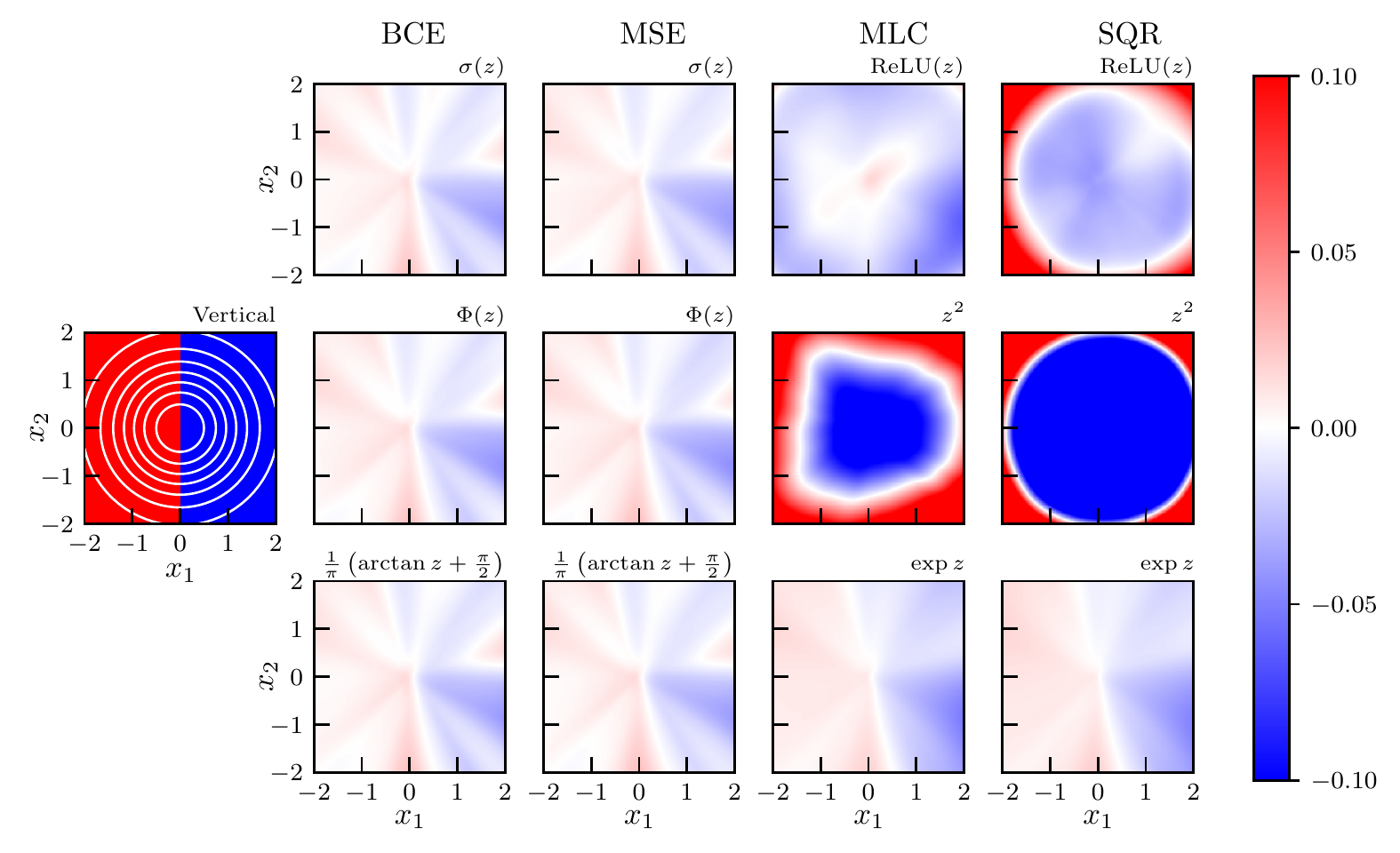}
    \includegraphics{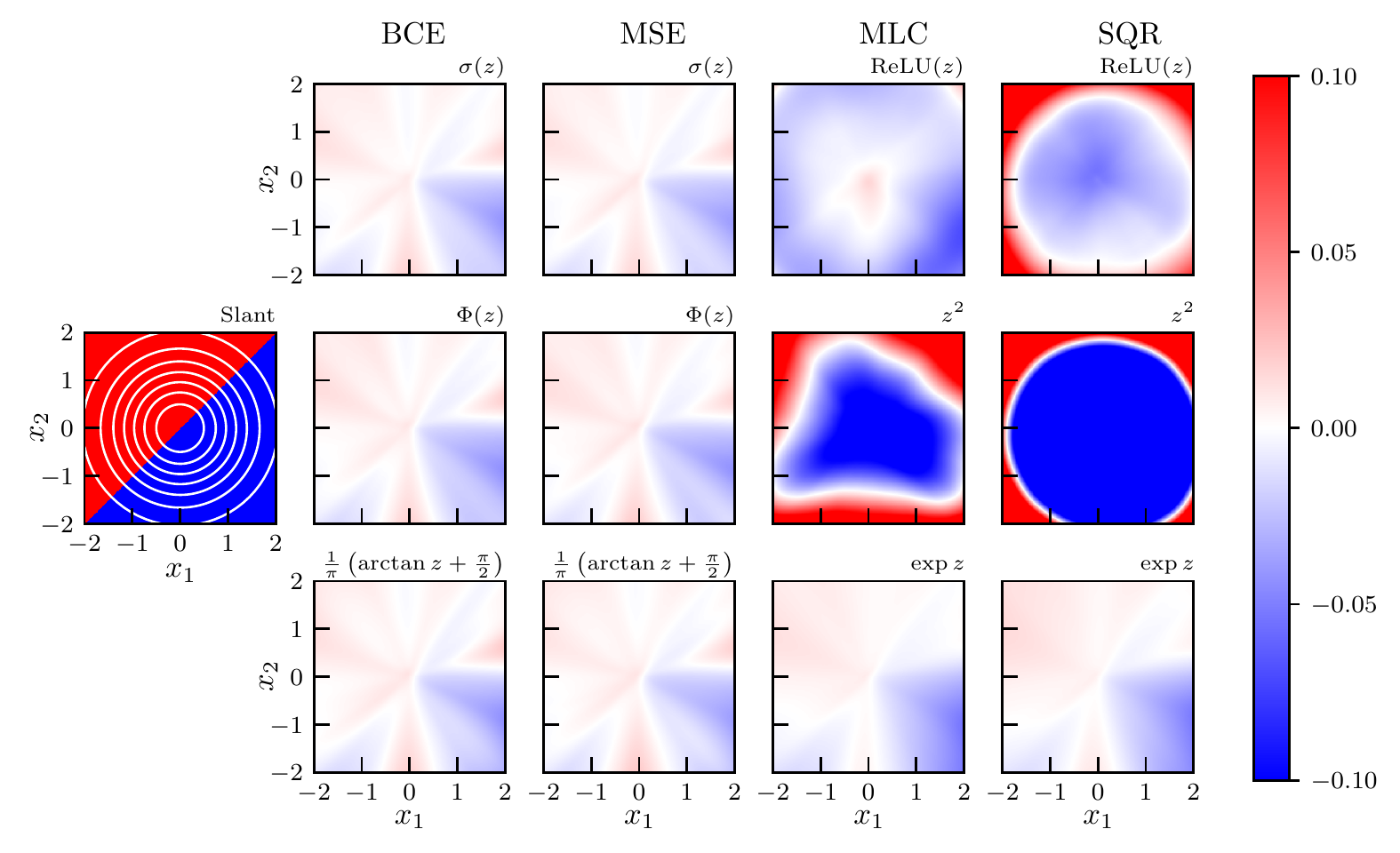}
    \caption{The likelihood ratio fits of the Vertical and Slant cases we examined in the multivariate Gaussians case studies. The first column plots the likelihood structure of each case; red regions are regions where \(\lr(x) \le 1\), and blue regions are regions where \(\lr(x) > 1\). Each row corresponds to a different loss functional, and each column corresponds to a different parametrization. The first three rows display the likelihood ratio fits for the Vertical case study and the second three rows display the likelihood ratio fits for the Slant case study. The plot is suggestively colored to show how the structure of the data corresponds to the structure in the likelihood ratio models.}
    \label{fig:mvn_vertical_slant}
\end{figure*}

\begin{figure*}[ht]
    \includegraphics{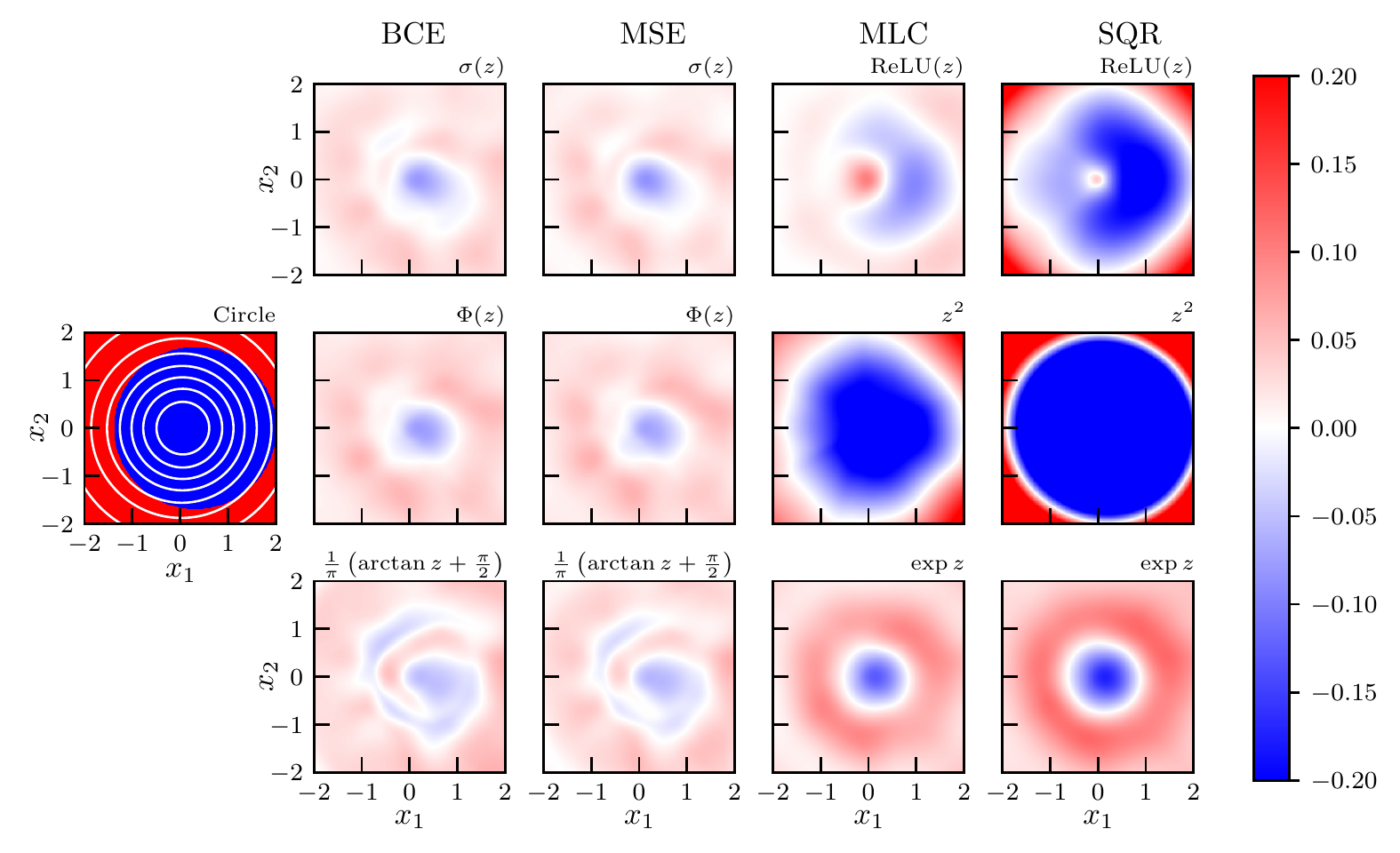}
    \includegraphics{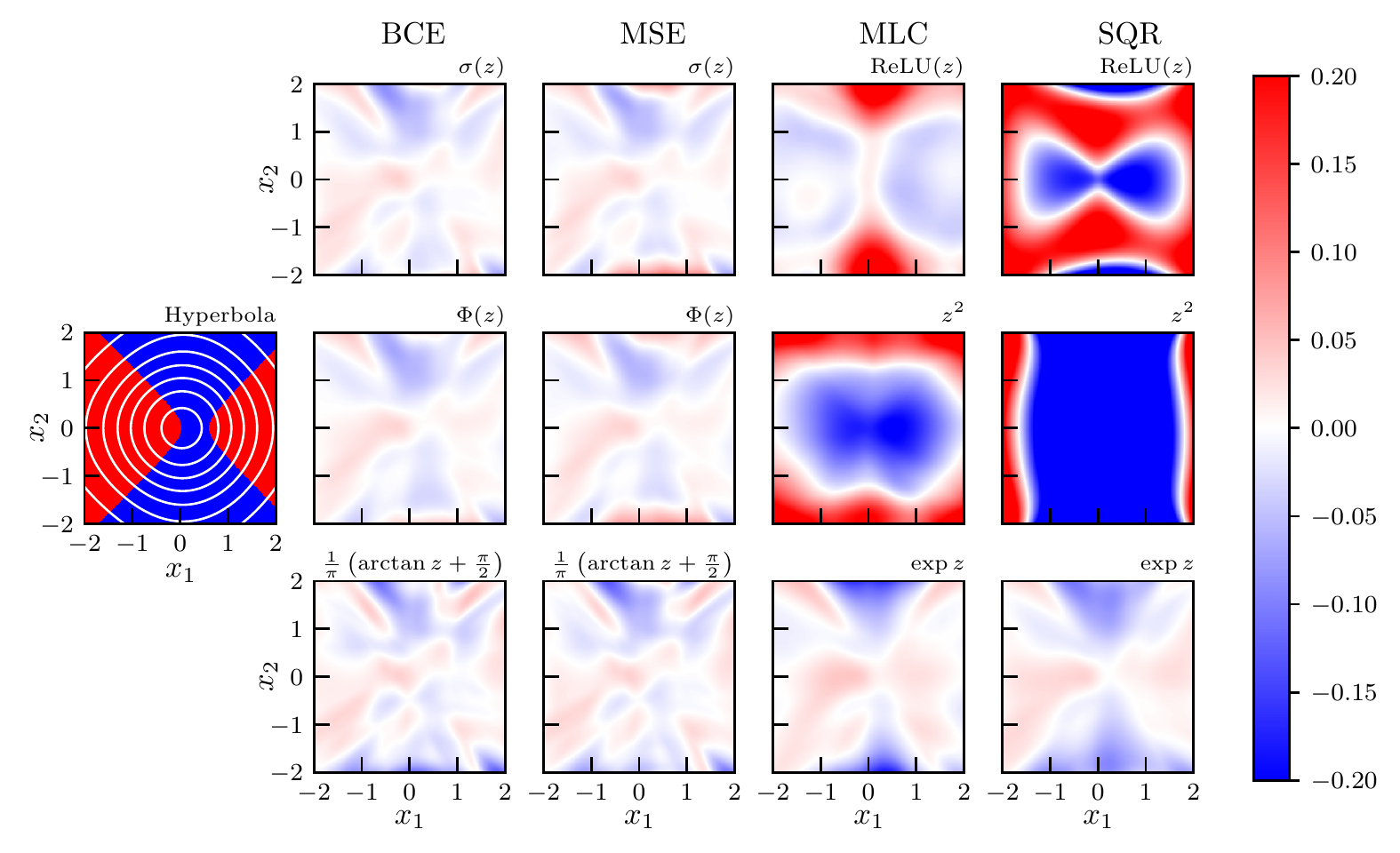}
    \caption{The likelihood ratio fits of the Circle and Hyperbola cases we examined in the multivariate Gaussians case studies, organized and colored as in \ref{fig:mvn_vertical_slant}.}
    \label{fig:mvn_circle_hyperbola}
\end{figure*}

\begin{figure*}
    \includegraphics{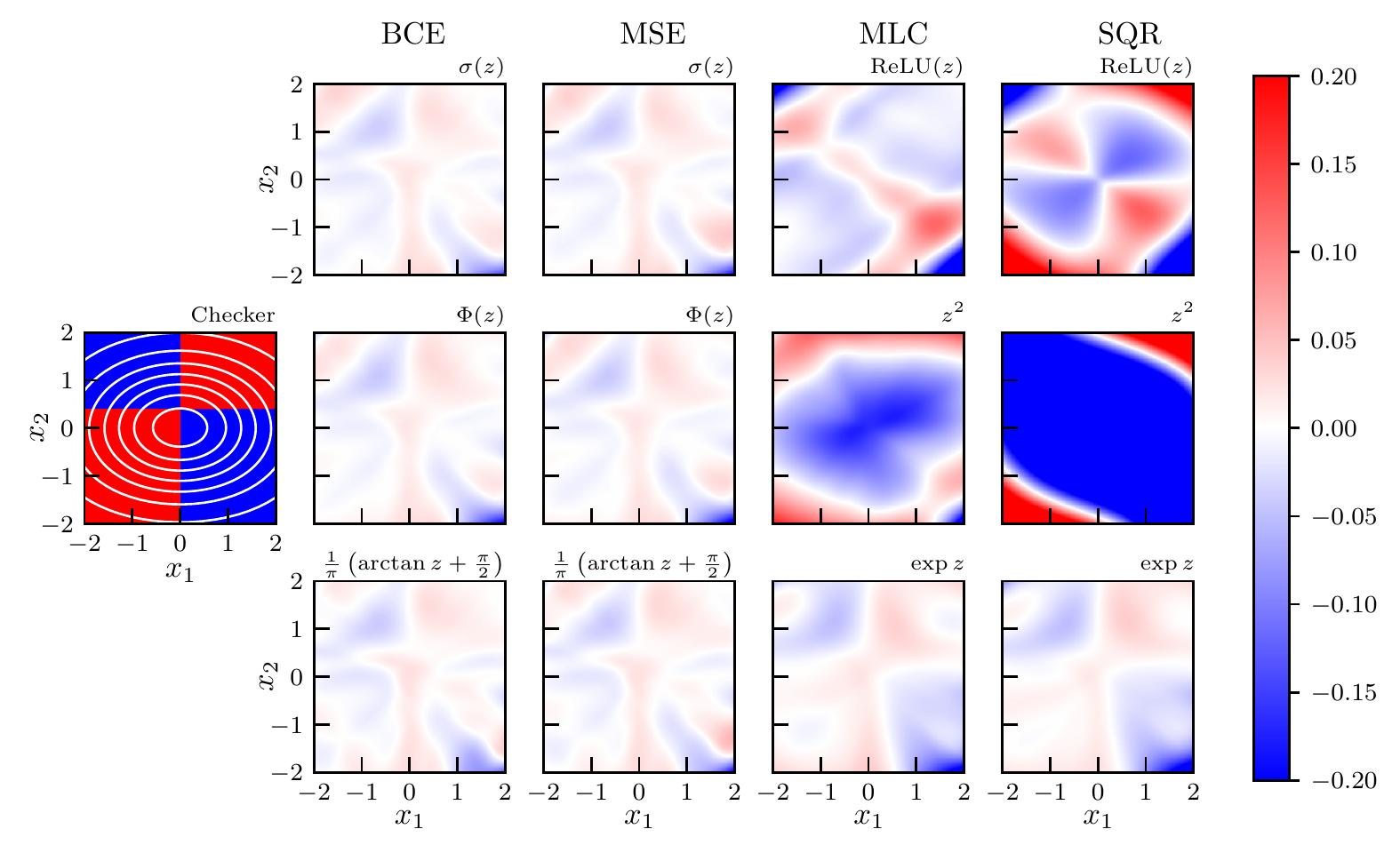}
    \caption{The likelihood ratio fits of the Checker case we examined in the multivariate Gaussians case studies, organized and colored as in \ref{fig:mvn_vertical_slant}.}
    \label{fig:mvn_checker}
\end{figure*}

The contour plots of the three different parametrizations for each of the four loss functionals for each of the five cases examined in the multivariate Gaussians case study are presented in Figures \ref{fig:mvn_vertical_slant}, \ref{fig:mvn_circle_hyperbola}, and \ref{fig:mvn_checker}.

\begin{figure*}
    \centering
    \includegraphics{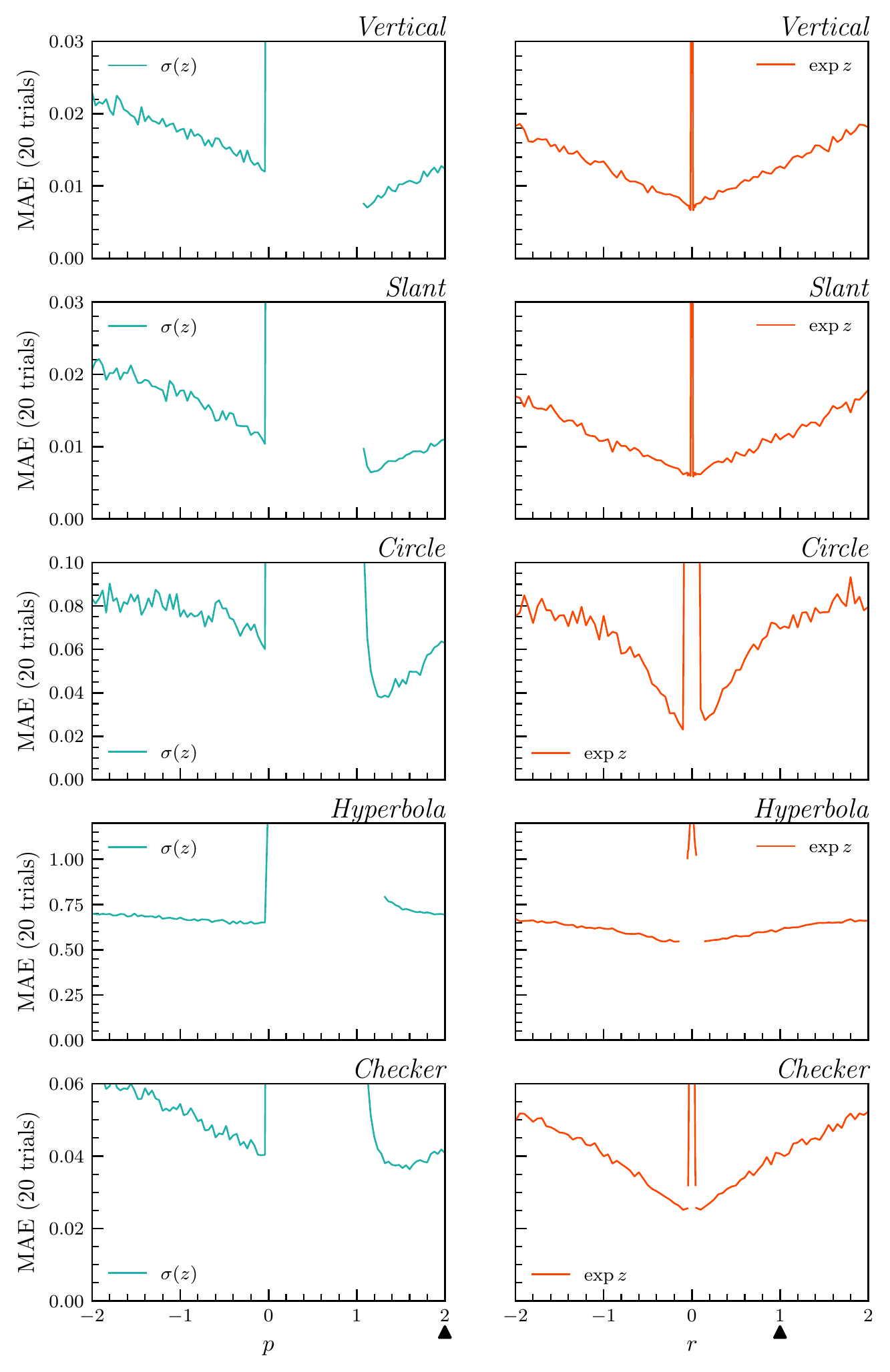}
    \caption{The mean absolute errors averaged over models trained on the generalized $p$-MSE and $r$-SQR loss families for the various multivariate Gaussian case studies. The classifiers were parametrized with the logistic and exponential activations for the $p$-MSE and $r$-SQR losses, respectively. The minimizing values of $p^*$ and $r^*$ are reported in Table \ref{tab:mvn_pr}. The arrows indicates the typical choice of $p=2$ and $r = 1$ for the MSE and SQR losses, respectively.}
    \label{fig:mvn_families}
\end{figure*}

Additionally, the plots of the MAEs of the $p$-MSE and $r$-SQR loss families are provided in Figure \ref{fig:mvn_families}.

\section{Comparison with BDTs}
\label{sec:bdts}
In this appendix, we compare the performance of a neural network (NN) likelihood ratio method with that of a gradient-boosted decision tree (BDT) method. While we parametrized the learned function $f$ as a neural network throughout all of our case studies, this is not a requirement. BDT methods generally are very well-performing, general-purpose, robust classifiers. 

In particular, we examined the performances of neural network and BDT methods as the dimension of the data on which the models were trained increased. In our first case study, we extended the Slant data from the multivariate Gaussians study to $d \in \{1, 2, 4, 8, 16, 32\}$ dimensions as follows:

\begin{align}
    X_0 &\sim \text{Normal}\left( \begingroup \renewcommand*{\arraystretch}{1.75} +\frac{1}{\sqrt{d}}\begin{bmatrix} 0.1 \\ \vdots \\ 0.1 \end{bmatrix}\endgroup, \mathbf{I}_d\right)\\
    X_1 &\sim \text{Normal}\left( \begingroup \renewcommand*{\arraystretch}{1.75} -\frac{1}{\sqrt{d}}\begin{bmatrix} 0.1 \\ \vdots \\ 0.1 \end{bmatrix}\endgroup, \mathbf{I}_d\right)
\end{align}

As in \ref{section:univariate_methods}, we trained 100 classifiers with $N \in \{10^2, 10^3, 10^4, 10^5, 10^6, 10^7\}$ samples from the Slant case. For each value of $N$, a random assortment of 75\% of the samples were used for training and the remaining 25\% were used for validation. A total of $100,000$ samples were used for estimating the MAE for each value of $N$.

The NN architectures were identical to those used throughout all of our previous studies, as was the training procedure and optimization. The BDT classifiers were implemented using \textsc{Scikit Learn} \cite{scikit-learn} with the default learning rate of $0.1$ and early stopping with a patience of 10. No detailed hyperparameter optimization was done.

\begin{figure*}[ht]
    \centering
    \includegraphics{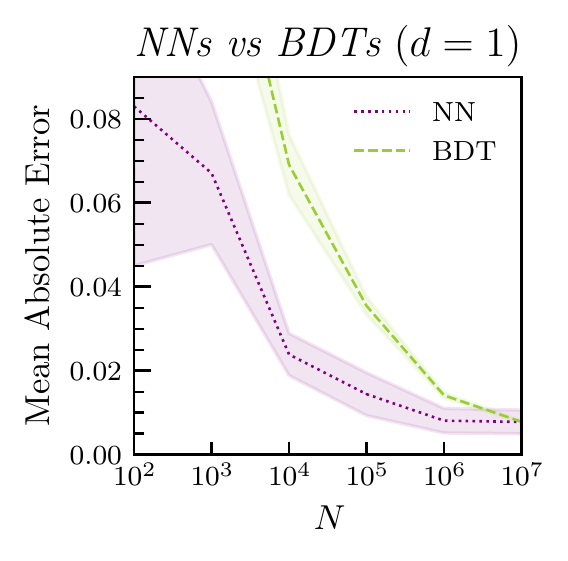} 
    \includegraphics{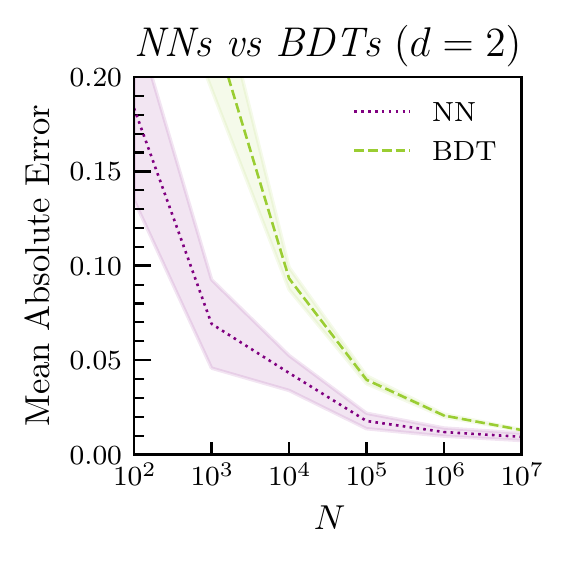} 
    \includegraphics{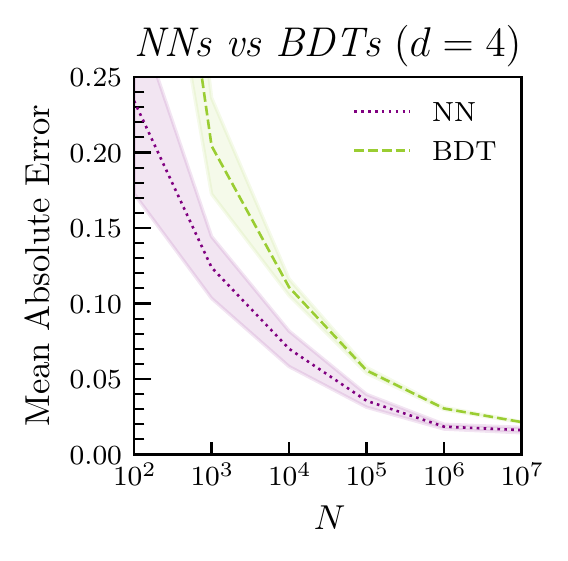} 
    \includegraphics{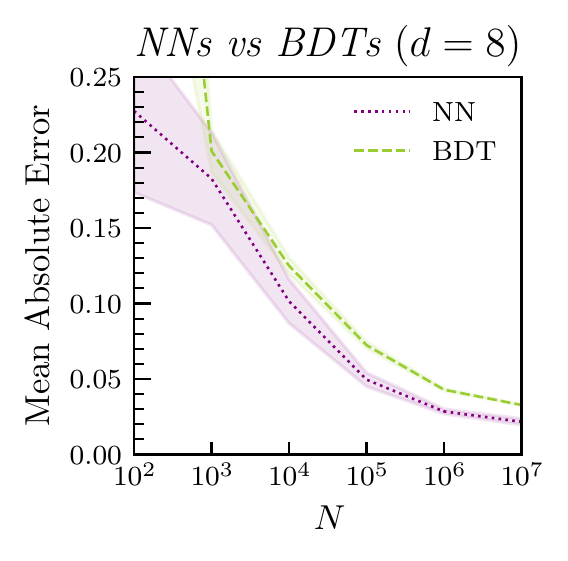}
    \includegraphics{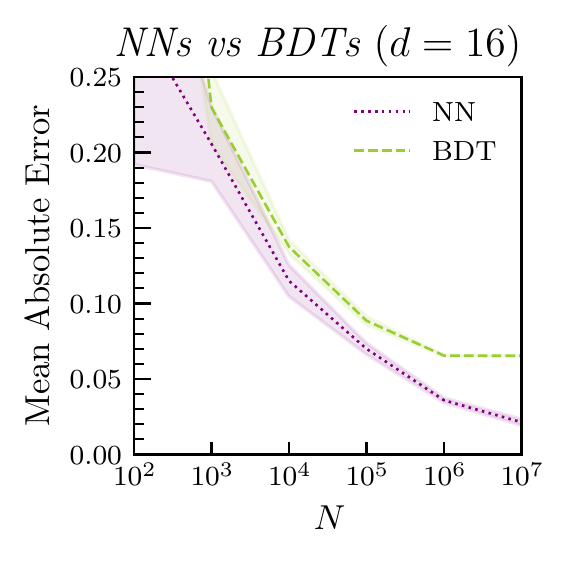} 
    \includegraphics{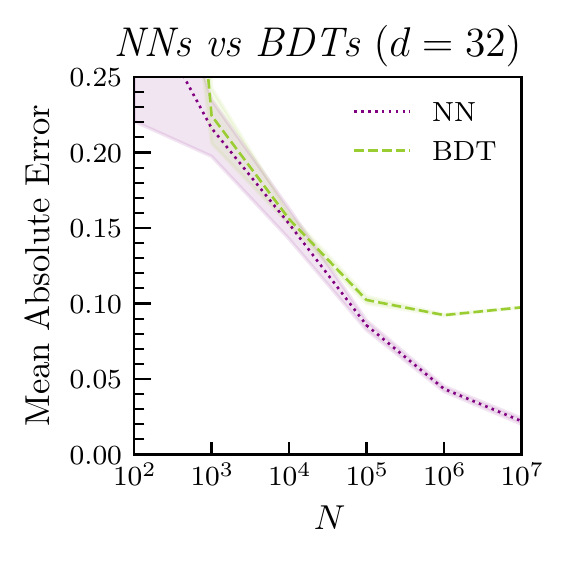} 
    \caption{The mean absolute errors of the neural network (NN) and boosted decision tree (BDT) classifiers for increasing sample size and dimension for the Slant classification problem. The mean absolute error of a classifier parametrization for each sample size was computed by averaging over the empirical mean absolute errors of 100 models. The shaded regions correspond to one standard deviation in the mean absolute errors of the 100 models.}
    \label{fig:trees_slant}
\end{figure*}

Figure \ref{fig:trees_slant} displays how the mean absolute error between the two classifier parametrizations compares as the sample size increases for the Slant dataset in increasingly higher dimensions. While the BDT tends to have a smaller variance than the NN, the NN always outperforms the BDT. Moreover, in higher dimensions, the NN performs significantly better than the BDT.

Next, we compared the two parametrizations' performances on the physics dataset from \ref{sec:hep}. The $d = 1$ dataset consisted of just the transverse momentum $p_T$, the $d = 2$ dataset consisted of the transverse momentum $p_T$ and the rapidity $y$, and the $d = 4$ dataset consisted of the transverse momentum $p_T$, the rapidity $y$, the azimuthal angle $\phi$, and the mass $m$. Each dataset was generated by training a Normalizing Flow \cite{normalizingflow} on the appropriate variables from the original physics dataset.

\begin{figure*}[ht]
    \centering
    \includegraphics{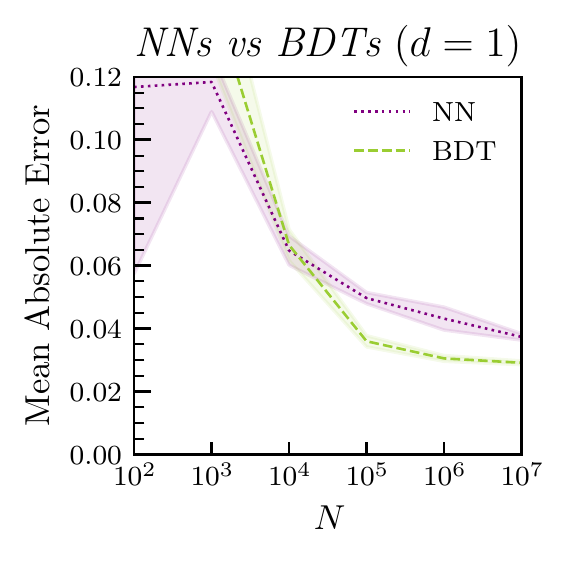} 
    \includegraphics{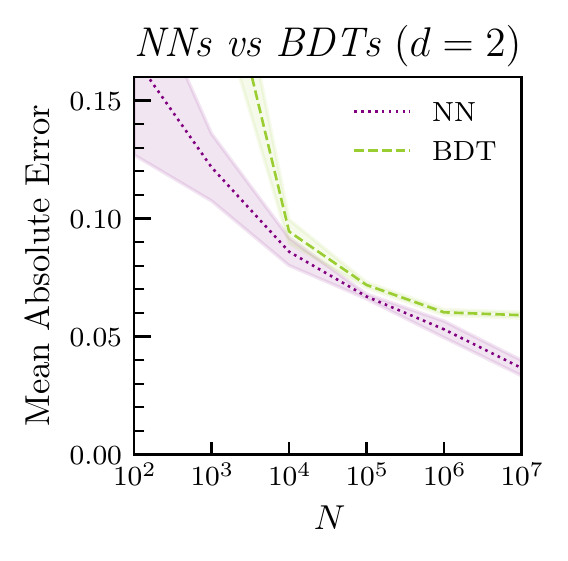} 
    \includegraphics{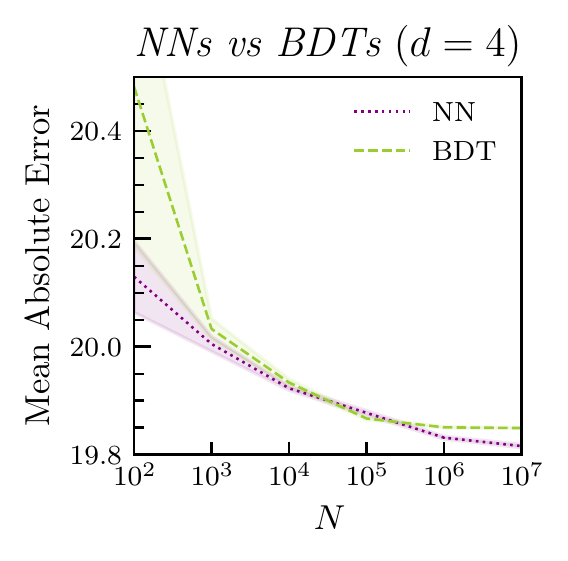}
    \caption{The mean absolute errors of the neural network and boosted decision tree classifiers for increasing sample size and dimension for the \textsc{Pythia}/\textsc{Herwig} + \textsc{Delphes} particle physics jet datasets. The mean absolute error of a classifier parametrization for each sample size was computed by averaging over the empirical mean absolute errors of 100 models. The shaded regions correspond to one standard deviation in the mean absolute errors of the 100 models.}
    \label{fig:trees_physics}
\end{figure*}

Once the datasets were generated, the methodology used was identical to the previous case study with the Slant dataset. Figure \ref{fig:trees_physics} displays how the two parametrizations compare against one another at different sample sizes and dimensions. We again see that the BDT is lower variance than the NN. Moreover, in this case, the BDT outperforms the NN at all sample sizes in the $d = 1$ case. However, in $d = 2$ and $d = 4$, the NN outperforms the BDT.

Across each of these two datasets, our results indicate that despite the generally strong performance of BDTs, NNs are still preferable in the context of the likelihood ratio trick.

\end{document}